\documentclass[traditabstract]{aa}
\bibpunct{(}{)}{;}{a}{}{,} 
\usepackage{graphicx}
\usepackage{mathrsfs}
\usepackage[varg]{txfonts}
\usepackage[T1]{fontenc}
\usepackage{color}
\usepackage{stfloats}
\usepackage[switch,pagewise]{lineno}
\DeclareRobustCommand{\rchi}{{\mathpalette\irchi\relax}}
          \newcommand{\irchi}[2]{\raisebox{\depth}{$#1\chi$}}

\begin{document}

\title{Microlensing of the broad emission line region in the lensed quasar J1004$+$4112}
\author{Damien Hutsem\'ekers\inst{1,}\thanks{Research Director F.R.S.-FNRS} 
   \and Dominique Sluse\inst{1}
   \and Đorđe Savić\inst{1}
   \and Gordon~T. Richards\inst{2}
       }
\institute{
    Institut d'Astrophysique et de G\'eophysique,
    Universit\'e de Li\`ege, All\'ee du 6 Ao\^ut 19c, B5c,
    4000 Li\`ege, Belgium
    \and
    Department of Physics, Drexel University, 32 S. 32nd Street, Philadelphia, PA 19104, USA
    }
%
%
%
%
\abstract{J1004+4112 is a lensed quasar for which the first broad emission line profile deformations due to microlensing were identified. Detailed interpretations of these features have nevertheless remained controversial. Based on 15 spectra obtained from 2003 to 2018, in this work, we revisit the microlensing effect that distorts the \ion{C}{iv} broad emission line profile in J1004+4112. We take advantage of recent measurements of the image macro-magnification ratios, along with the fact that at one epoch, image B was not microlensed, thus constituting a reference spectrum to unambiguously characterize the microlensing effect observed in image~A. After disentangling the microlensing in images A and B, we show that the microlensing-induced line profile distortions in image~A, although variable, are remarkably similar over a period of 15 years. We find they are characterized by a strong magnification of the blue part of the line profile, a strong demagnification of the red part of the line profile, and a small-to-negligible demagnification of the line core. We used the microlensing effect, characterized by either the full magnification profile of the \ion{C}{iv} emission line or a set of four integrated indices, to constrain the broad emission-line region (BLR) size, geometry, and kinematics. For this purpose, we modeled the deformation of the emission lines considering three simple, representative BLR models: a Keplerian disk, an equatorial wind, and a biconical polar wind, with various inclinations with respect to the line of sight. We find that the observed magnification profile of the \ion{C}{iv} emission line in J1004+4112 can be reproduced with the simple BLR models we considered, without the need for more complex BLR features.  The magnification appears dominated by the position of the BLR with respect to the caustic network -- and not by the velocity-dependent size of the BLR. The favored models for the \ion{C}{iv} BLR are either the Keplerian disk or the equatorial wind, depending on the orientation of the BLR axis with respect to the caustic network. We also find that the polar wind model can be discarded. We measured the \ion{C}{iv} BLR half-light radius as $r_{1/2} = 2.8^{+2.0}_{-1.7}$ light-days.  This value is smaller than the BLR radius expected from the radius-luminosity relation derived from reverberation mapping, but it is still in reasonable agreement given the large uncertainties.
}
\maketitle
%
%
%

\section{Introduction}
\label{sec:intro}

\begin{table*}[h]
\caption{List of spectra used in this paper}
\label{tab:data}
\centering
\begin{tabular}{ccccccc}
\hline\hline
 \multicolumn{2}{c}{Date}  & Telescope & Spectral Range & $\Delta \lambda$ & Images & Reference\\
 (yyyymmdd) &    (MJD) & & (\AA) & (\AA) & &  \\
\hline
2003 05 31  &    52790  &     Keck I   & 3028-9700 & 6.5-7.1 &    ABCD   &    1,2,3   \\
2003 11 21  &    52964  &     ARC 3.5m & 3890-9350 & 8.4  &    AB     &    3   \\
2003 11 30  &    52973  &     ARC 3.5m & 3890-9350 & 8.4  &    AB     &    3   \\
2003 12 01  &    52974  &     ARC 3.5m & 3890-9350 & 8.4  &    AB     &    3   \\
2003 12 22  &    52995  &     ARC 3.5m & 3890-9350 & 8.4  &    AB     &    3   \\
2004 03 26  &    53090  &     ARC 3.5m & 3890-9350 & 8.4  &    AB     &    4,5   \\
2004 04 10  &    53105  &     ARC 3.5m & 3890-9350 & 8.4  &    AB     &    6   \\
2004 04 26  &    53121  &     ARC 3.5m & 3890-9350 & 8.4  &    AB     &    6   \\
2004 05 13  &    53138  &     ARC 3.5m & 3890-9350 & 8.4  &    AB     &    6   \\
2004 05 28  &    53153  &     ARC 3.5m & 3890-9350 & 8.4  &    AB     &    6   \\
2004 12 08  &    53347  &     ARC 3.5m & 3890-9350 & 8.4  &    AB     &    6   \\
2004 12 17  &    53356  &     ARC 3.5m & 3890-9350 & 8.4  &    AB     &    6   \\
2006 05 01  &    53856  &     ARC 3.5m & 3890-9350 & 8.4  &    AB     &    6   \\ 
2008 10 17  &    54756  &     SAO 6m   & 3650-7540 &  10  &    ABCD   &    7   \\ 
2018 02 07  &    58156  &     SAO 6m   & 3650-7250 &   5  &    ABCD   &    7   \\
\hline
\end{tabular}
\tablebib{(1)~\citet{2003Inada}; (2)~\citet{2004Oguri}; (3)~\citet{2004Richards}; (4) \citet{2004bRichards};
          (5) \citet{2006Gomez}; (6) This paper; (7)~\citet{2020Popovic}}
\end{table*}

The quasar J1004+4112 (SDSS~J100434.80$+$411239.2 for component A) at redshift $z$ = 1.734 is a gravitationally lensed quasar with four images separated by angular distances between 3.7$\arcsec$ and 14.6$\arcsec$, with the lens being a galaxy cluster at $z$ = 0.68 \citep{2003Inada,2004Oguri}. Although the four components have the same redshift and display similar spectra, there are significant differences  between the broad emission-line profiles observed in the different images. In particular, image~A exhibits a strong, time-dependent, blue emission-line wing enhancement in the high-ionization lines, interpreted with a varying microlensing magnification of the broad emission-line region (BLR), rapid spectral changes intrinsic to the quasar and seen with a time delay between the different images being excluded \citep{2004Richards}. However, the absence of apparent continuum microlensing in image A and the recurrence of some line profile deformations have been considered as a challenge for the microlensing interpretation \citep{2006Gomez}. \citet{2006Lamer} proposed a mixed scenario involving a stationary microlensing effect, with the time-dependent changes being due to intrinsic variability rather than to a moving caustic, under the assumption that image B was not microlensed itself. Based on X-ray spectra, these authors also ruled out the possibility that the line profile differences observed between the four images are due to different lines of sight through quasar outflows that lead to differential absorption \citep{2006Green}. In parallel, theoretical simulations demonstrated that the microlensing of a biconical BLR can explain the observations \citep{2007Abajas}. More recent observations \citep{2012Motta,2020Popovic} showed that the blue emission-line wing enhancement in component A, although variable in strength, is a long-term effect. With the evidence of chromatic microlensing of the continuum \citep{2012Motta}, these observations provided additional support to the microlensing interpretation.

In the present paper, we revisit the microlensing effect in J1004+4112 using spectra obtained over 15 years including previously unpublished ones. Our analysis takes advantage of the fact that at one epoch, image B was not affected by microlensing and thus may be used as a reference to unambiguously characterize the microlensing effect at work in image A. It also benefits from recent radio measurements that provide the macro-magnification ratios of the different components \citep{2021Hartley}. The data are presented in Sect.~\ref{sec:data}. The microlensing-induced deformations of the \ion{C}{iv} broad emission line profile are discussed and quantified in Sect.~\ref{sec:description}. In order to characterize the BLR geometry and kinematics, and estimate its size, as was recently done for other lensed quasars \citep{2019Hutsemekers,2021Hutsemekers}, we compare the observed \ion{C}{iv} line profile deformations to microlensing simulations based on simple, representative BLR models magnified by a caustic network. The method is described in Sect.~\ref{sec:modeling} and the results are given in Sect.~\ref{sec:results}. Our conclusions are given in the final section.

\section{Data}
\label{sec:data}

For our analysis, we used a series of 12 spectra obtained from November 2003 to May 2006 with the ARC 3.5m telescope at Apache Point Observatory. The spectra of images A and B were secured using the same instrument setup and data reduction for each epoch. Details on this data set are given in \citet{2004Richards}, with an analysis of the 2003 subsample.

We also considered the higher quality spectra of images A, B, C, and D, obtained with the Keck I telescope in May 2003. These spectra are described in \citet{2003Inada}, \citet{2004Oguri}, and \citet{2004Richards}.

Two series of good quality spectra of the four images obtained in 2008 and 2018 with the 6m telescope of the Special Astrophysical Observatory (SAO) were also added to our sample. These spectra are described in \citet{2020Popovic} and publicly available at the Strasbourg astronomical Data Center (CDS).

Table~\ref{tab:data} lists the main characteristics of the considered spectra; $\Delta\lambda$ is the spectral resolution. In total, there are 15 spectra unevenly distributed between 2003 and 2018. In the following, we focus on the \ion{C}{iv} broad emission line which is strong, unblended with other emission lines over most of its profile, and present in all spectra.

\begin{figure}
\resizebox{0.95\hsize}{!}{\includegraphics*{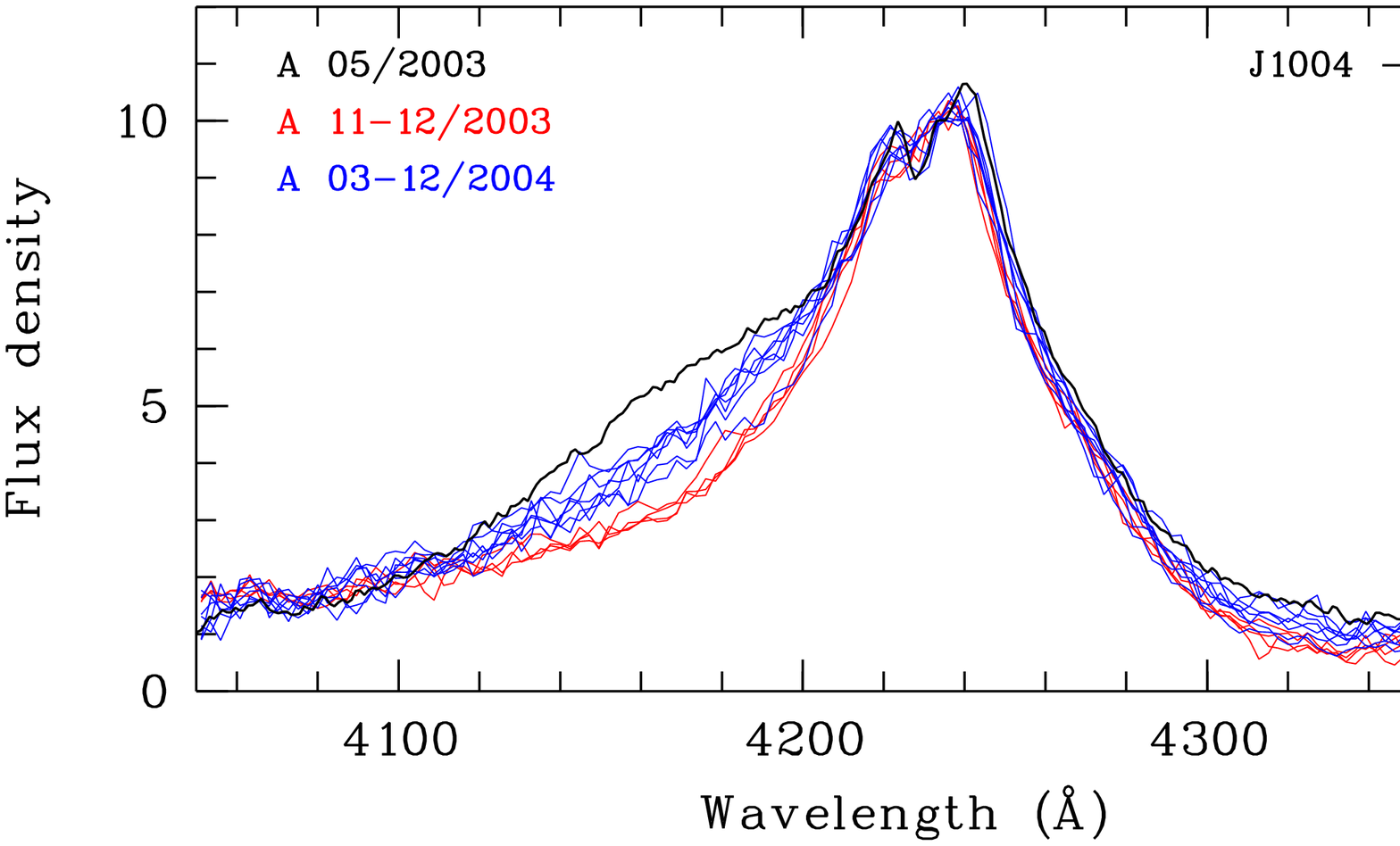}}\\
\resizebox{0.95\hsize}{!}{\includegraphics*{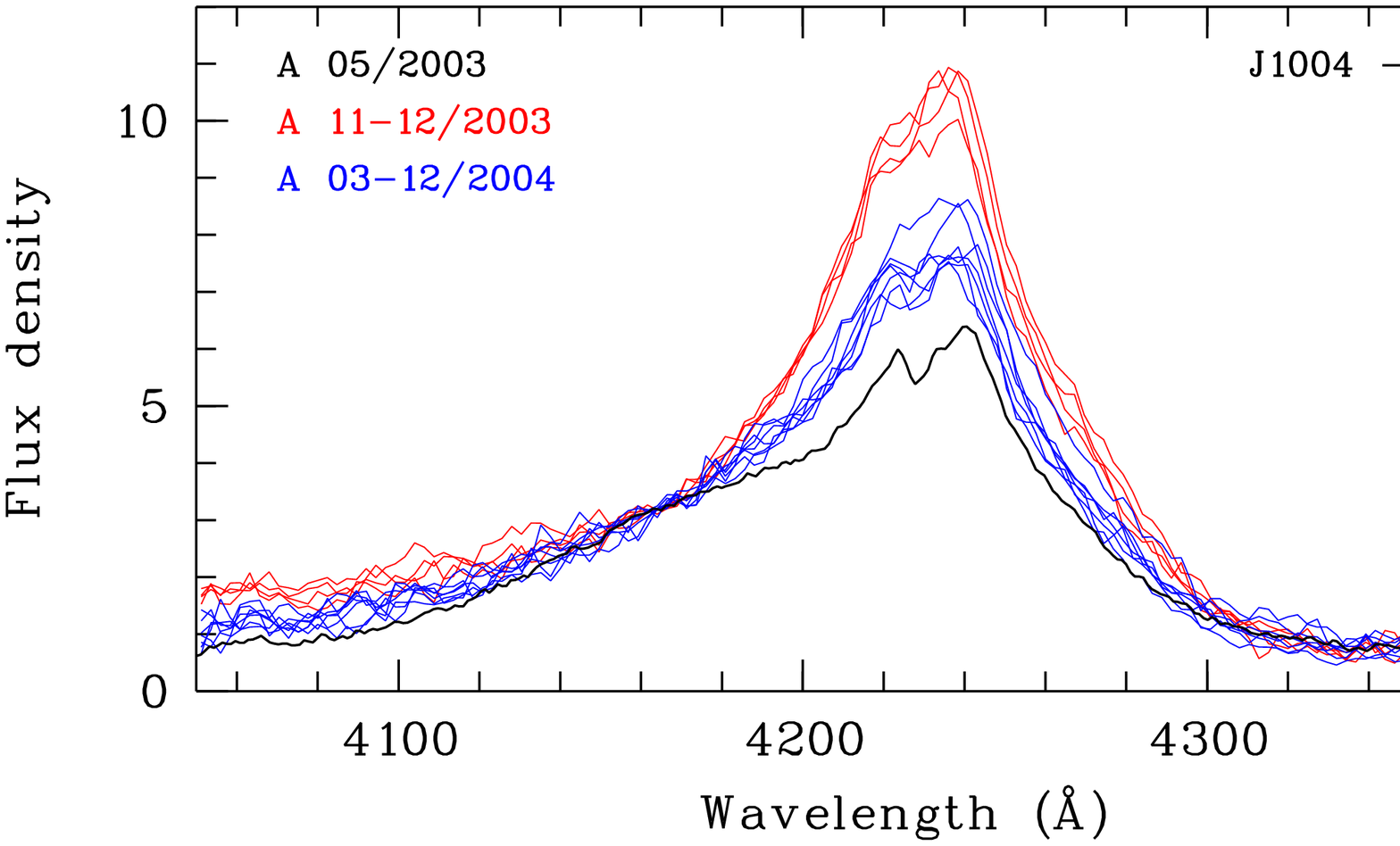}}\\
\resizebox{0.95\hsize}{!}{\includegraphics*{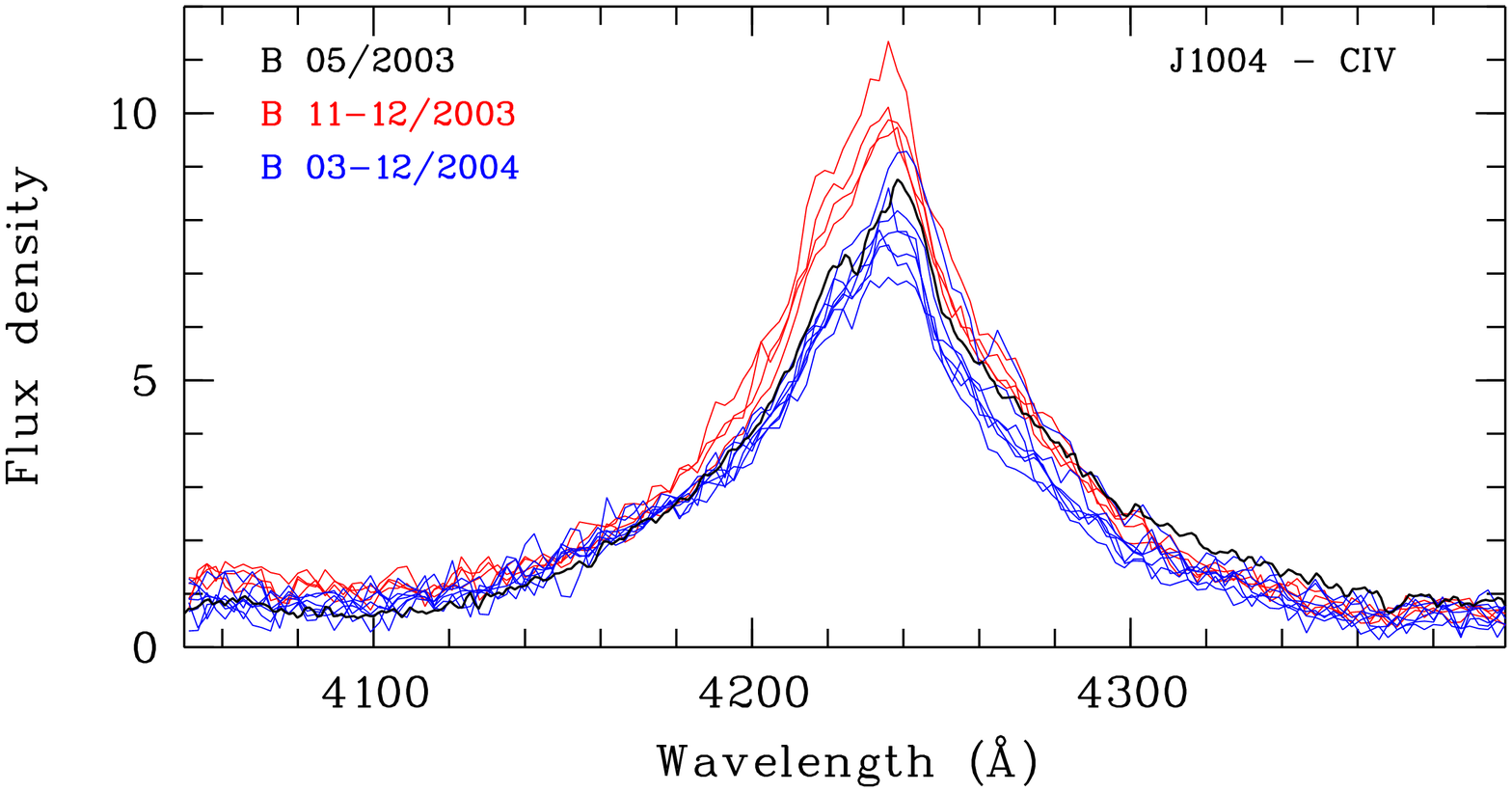}}\\
\caption{Continuum-subtracted \ion{C}{iv} emission line profiles observed in 2003 and 2004. {\bf Top:} Profiles in image A normalized to the peak.  {\bf Middle:} Profiles in image A normalized to the blue wing.  {\bf Bottom:} Profiles in image B normalized to the blue wing.}
\label{fig:profiles}
\end{figure}

\section{Broad emission-line microlensing in J1004+4112}
\label{sec:description}

\begin{figure*}[t]
\resizebox{\hsize}{!}{\includegraphics*{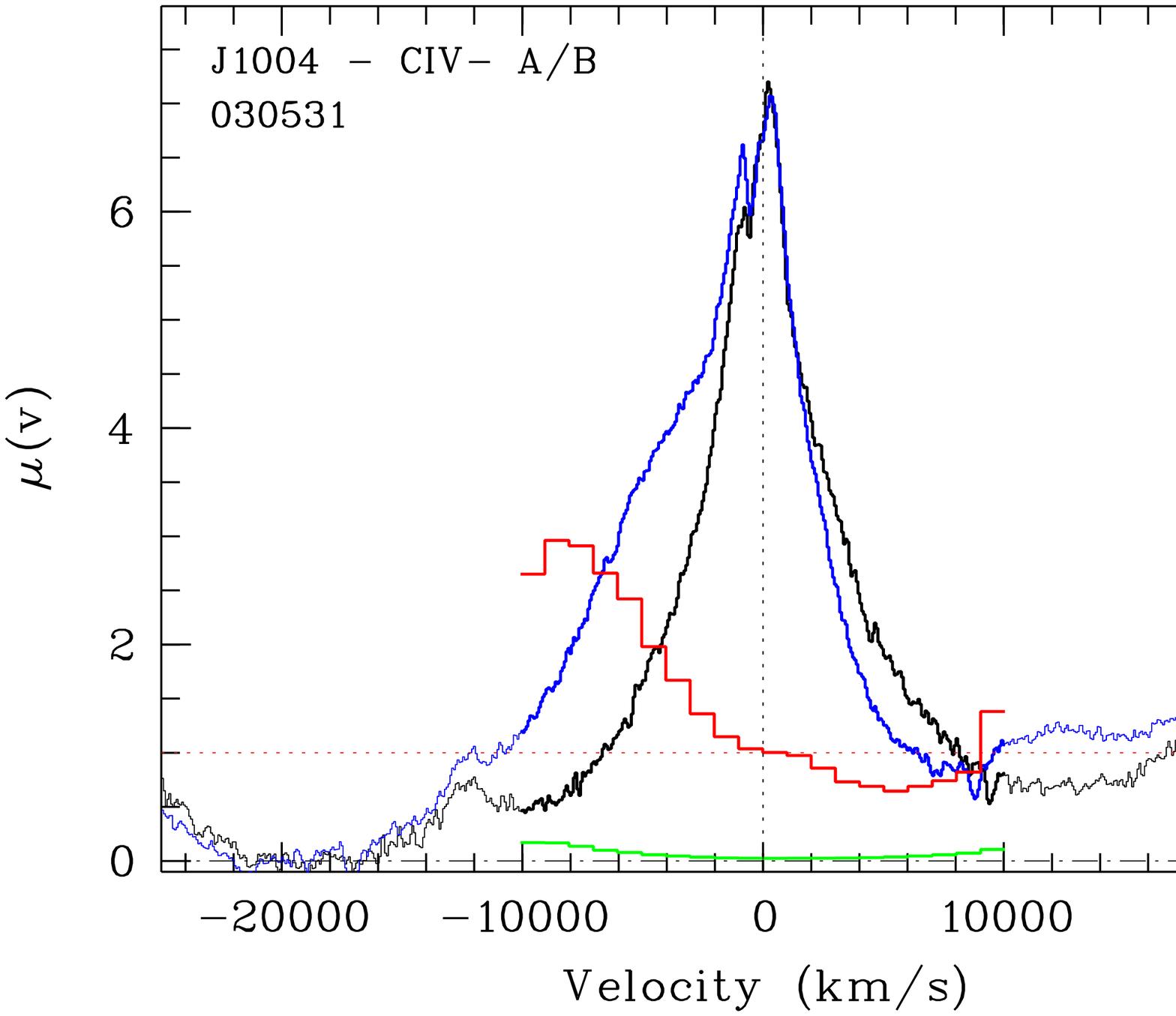}\includegraphics*{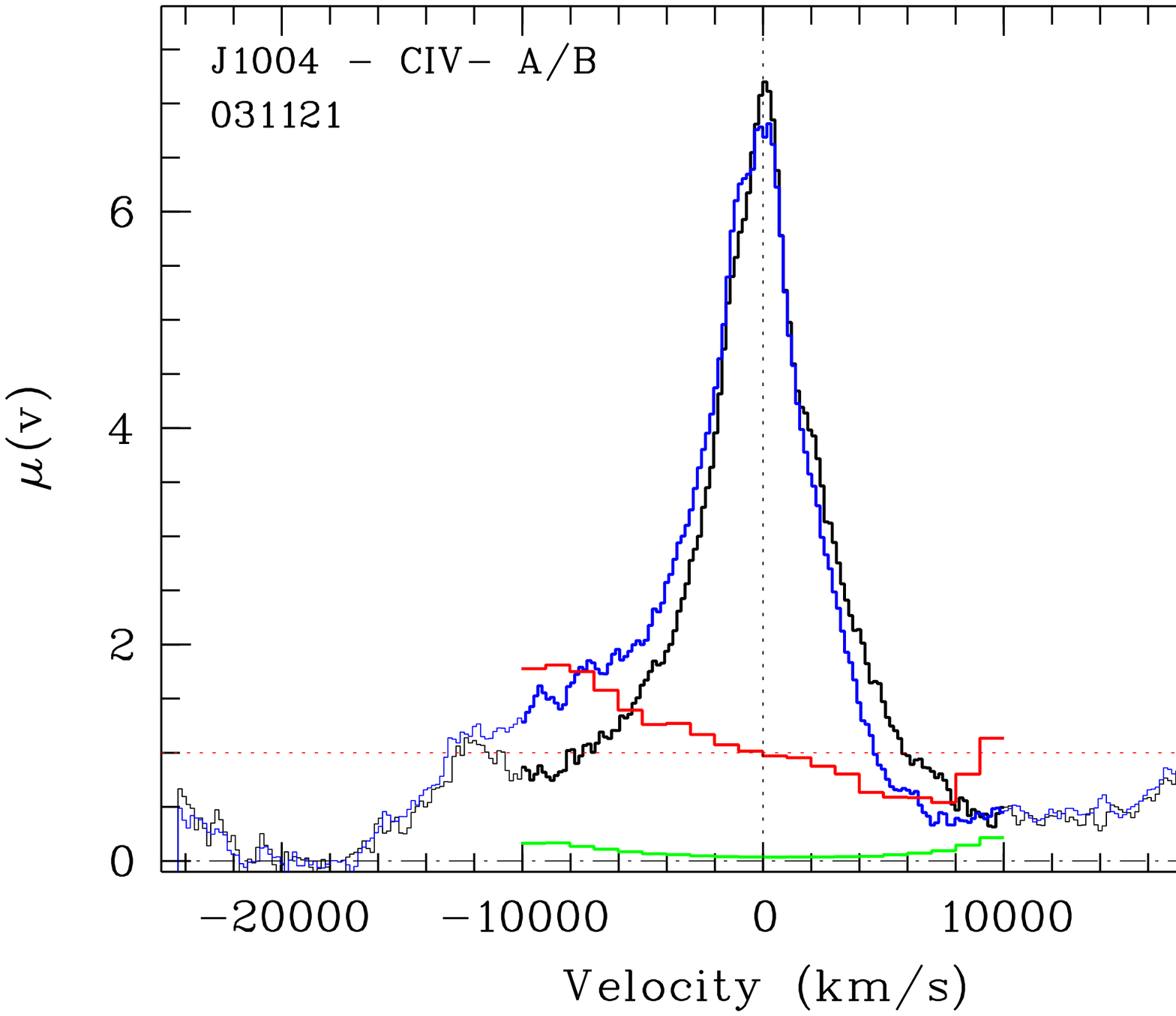}}\\
\resizebox{\hsize}{!}{\includegraphics*{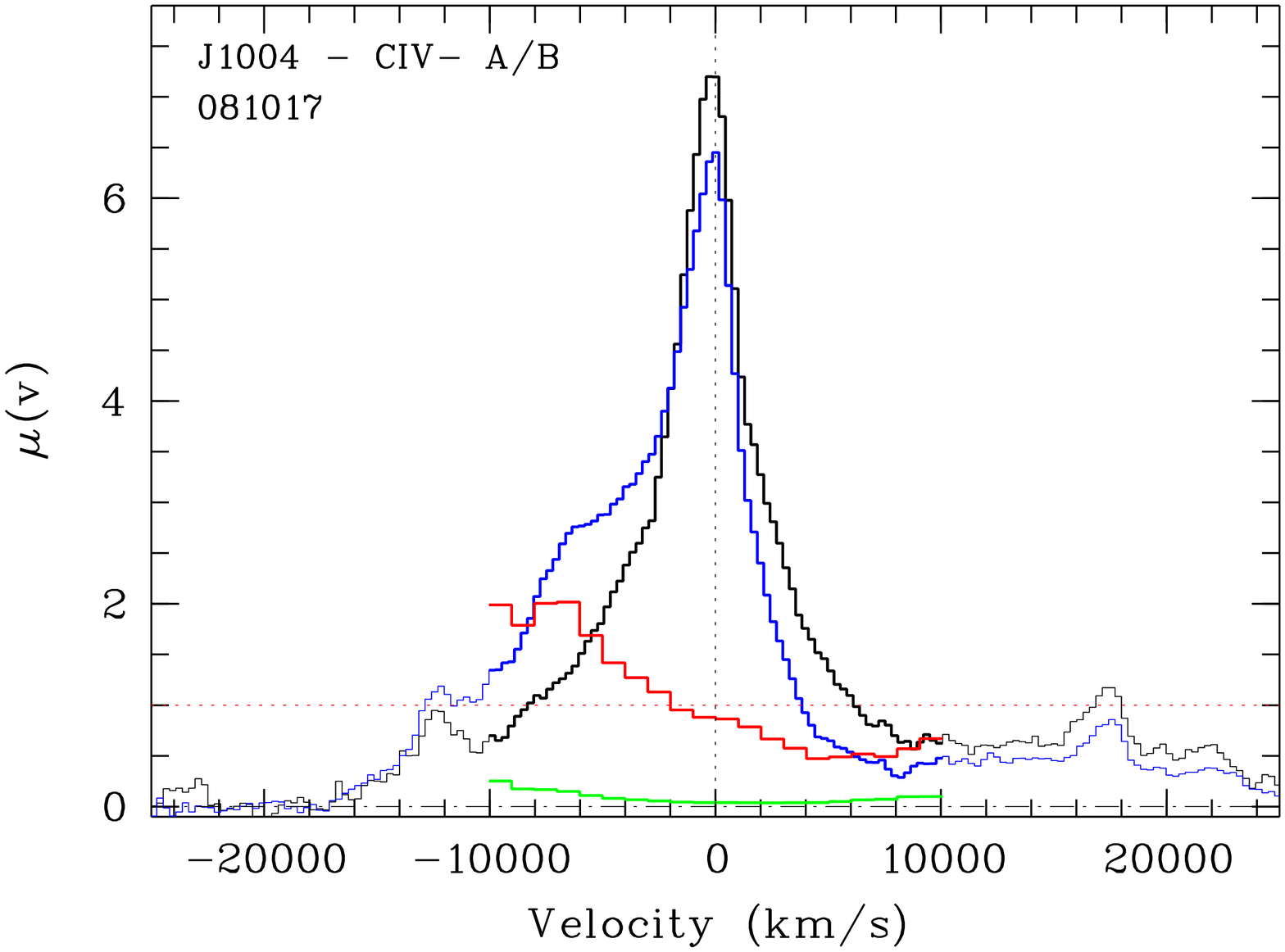}\includegraphics*{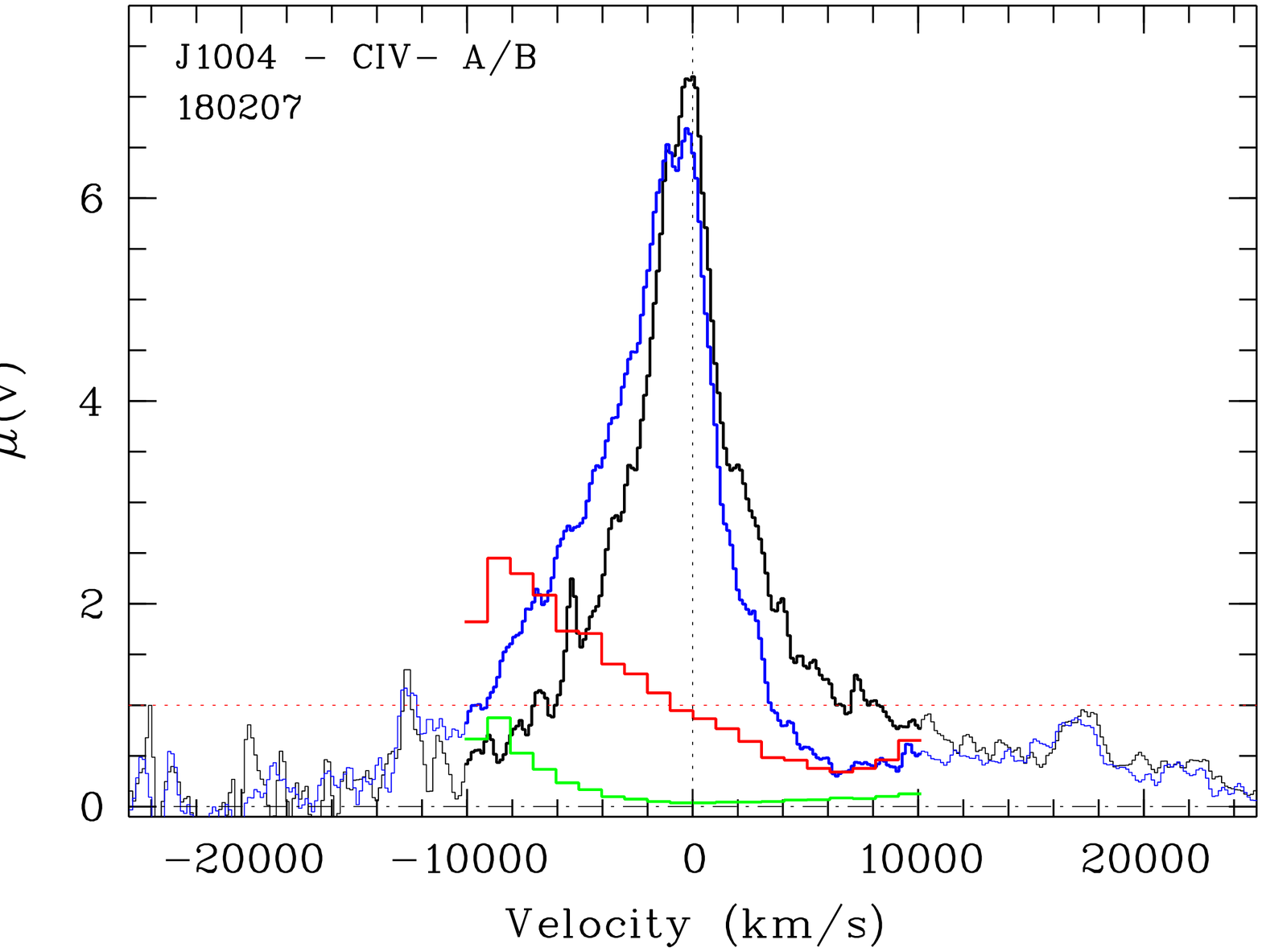}}
\caption{Examples of $\mu(v)$ magnification profiles of \ion{C}{iv} at four epochs (in red) computed from spectra of images A and B that were simultaneously recorded. The $\mu(v)$ profiles are binned into 20 spectral elements.  The superimposed line profiles from image A (in blue) and B (in black) are continuum-subtracted, corrected for the macro-magnification ratio, and arbitrarily rescaled for convenience. The thicker line indicates the part used in the computation of $\mu(v)$. The error of $\mu(v)$ is shown in green. The observation date is given as yymmdd. The zero-velocity corresponds to the \ion{C}{iv} $\lambda$1549 wavelength at the redshift $z$ = 1.734.}
\label{fig:exmuv}
\end{figure*}

\begin{figure*}[t]
\resizebox{\hsize}{!}{\includegraphics*{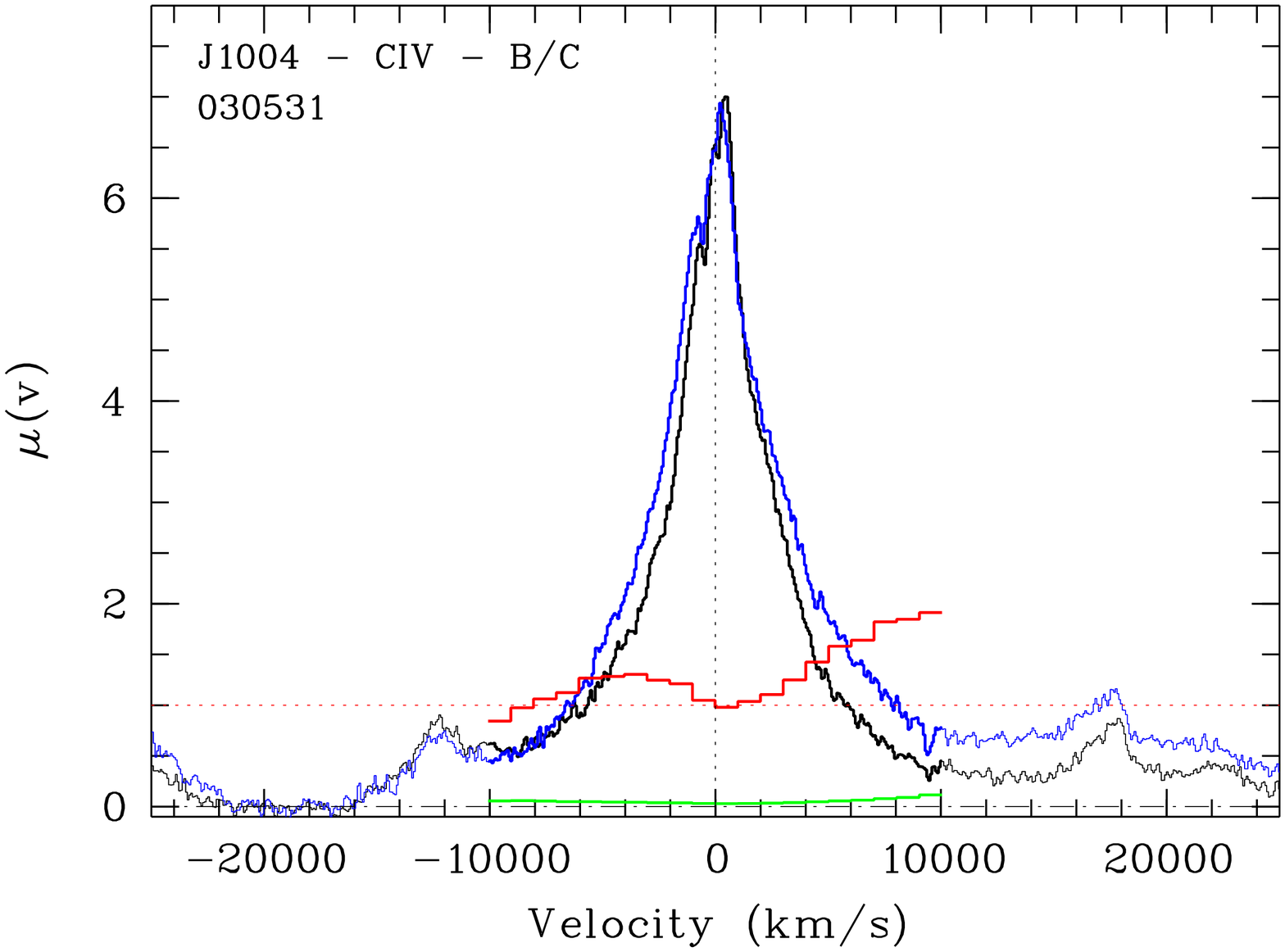}\includegraphics*{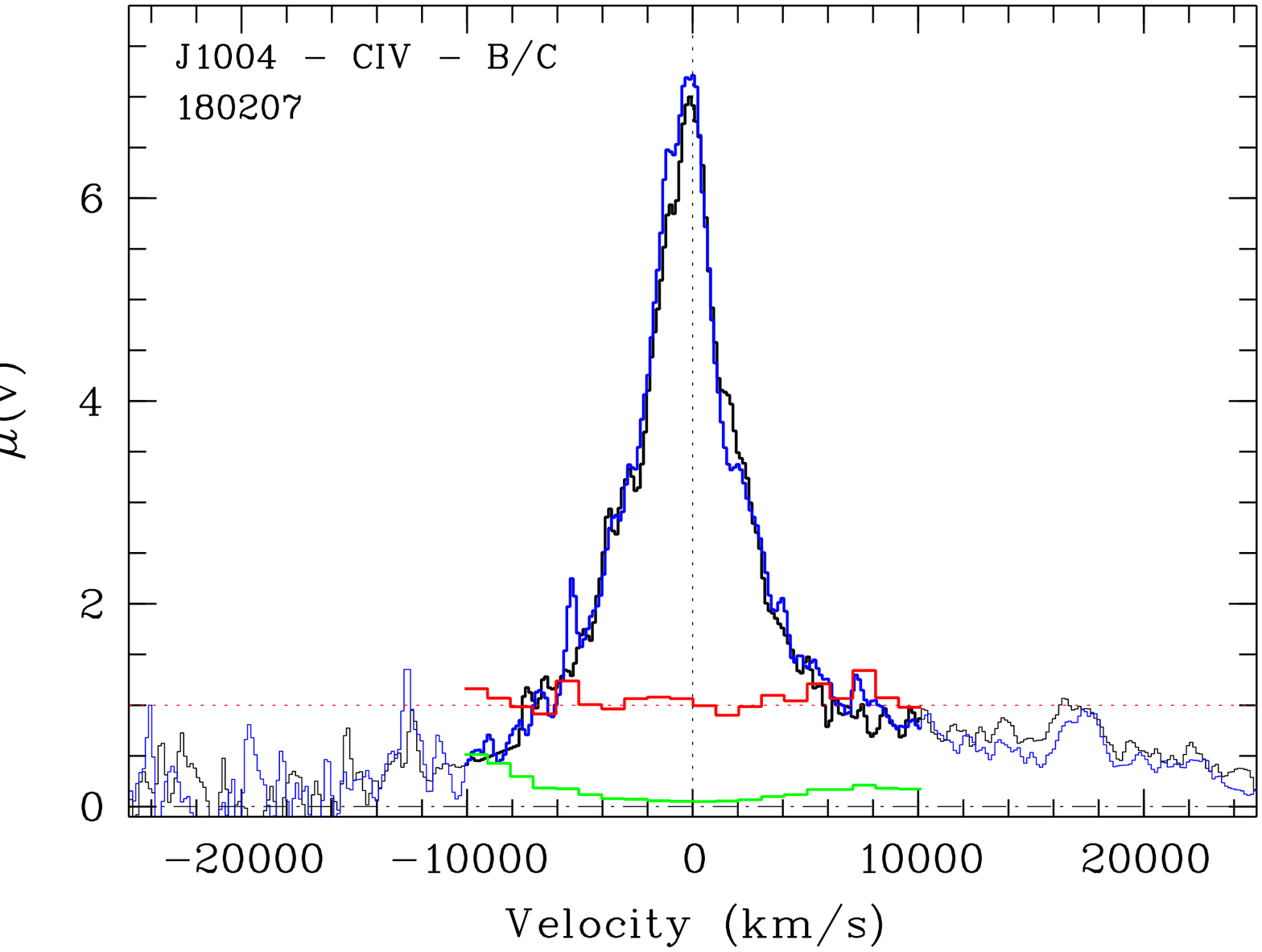}}
\caption{$\mu(v)$ magnification profile of \ion{C}{iv} (in red) computed from the spectra of images B and C obtained in May 2003 and February 2018, and binned into 20 spectral elements, as in Fig.~\ref{fig:exmuv}.  The line profiles from image B (in blue) and C (in black) are continuum-subtracted,  corrected for the macro-magnification ratio and arbitrarily rescaled for convenience. The error of $\mu(v)$ is shown in green.}
\label{fig:muvbc}
\end{figure*}

\begin{figure}[t]
\resizebox{\hsize}{!}{\includegraphics*{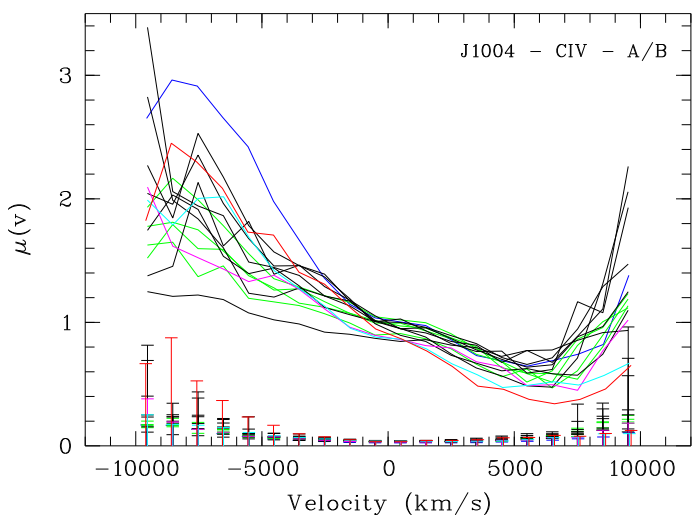}}
\caption{Time series of $\mu(v)$ profiles of \ion{C}{iv} computed from the spectra of images A and B obtained at the same epoch. Four epochs are emphasized: May 2003 (blue), May 2006 (cyan), October 2008 (magenta), and February 2018 (red); the other $\mu(v)$ profiles computed from spectra obtained in 2003 and 2004 are shown in green and black, respectively. The half error bars of $\mu(v)$ are drawn from zero for clarity.}
\label{fig:muvall}
\end{figure}

\begin{figure*}[t]
\resizebox{\hsize}{!}{\includegraphics*{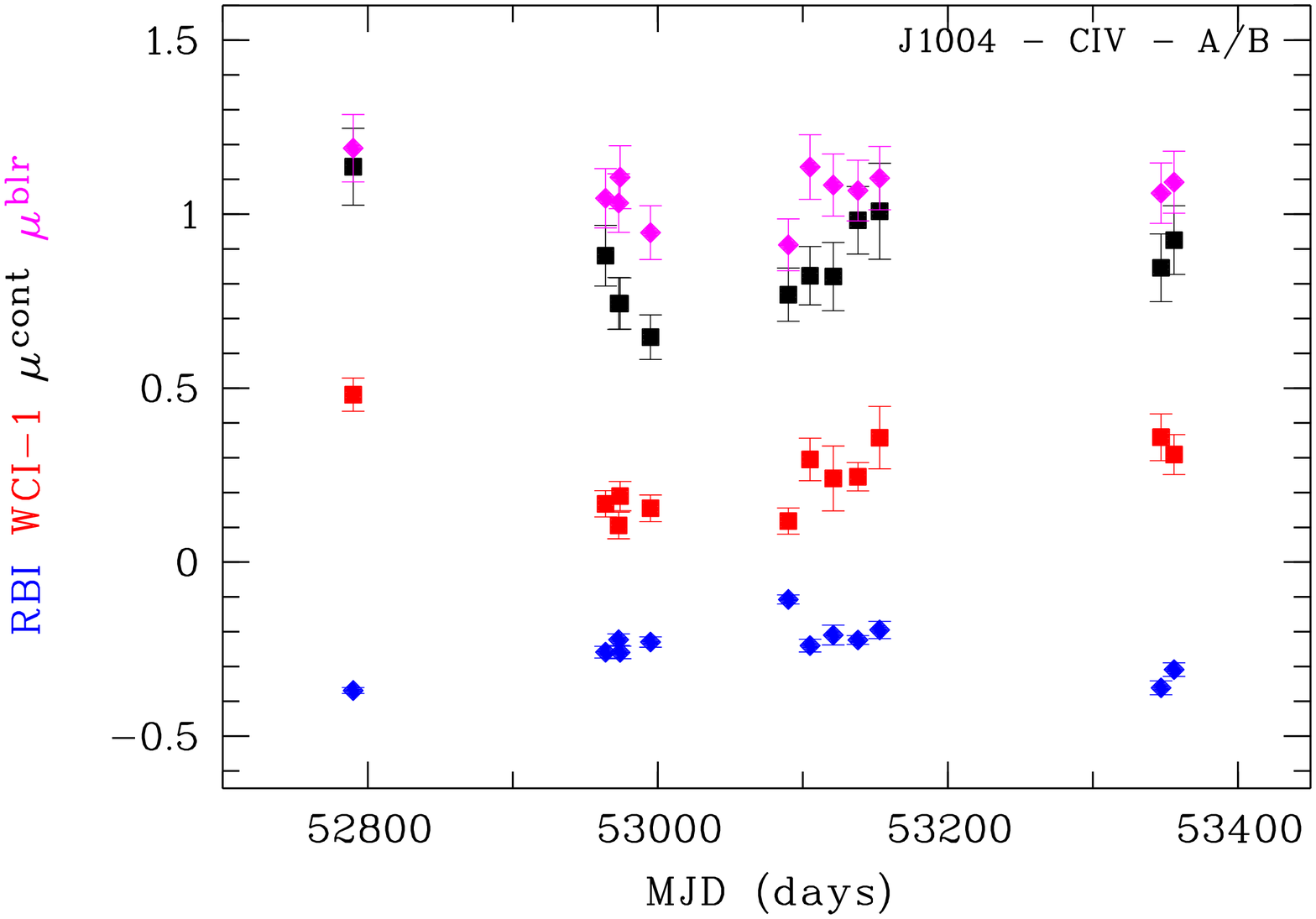}\includegraphics*{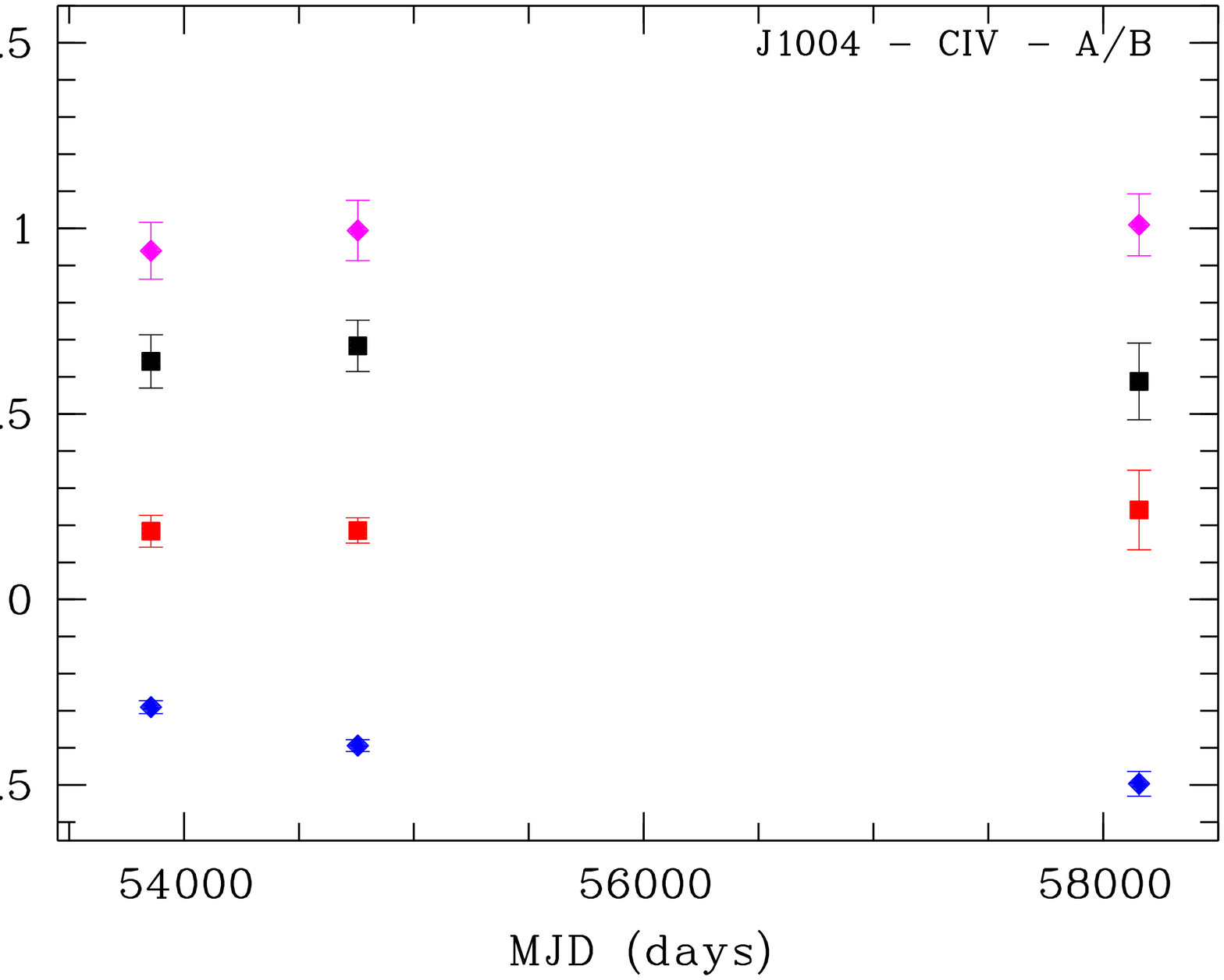}}\\
\caption{Time series of the four indices computed from the spectra of images A and B obtained at the same epoch. The two panels have different scales for the abscissae. At all epochs, $WCI > 1$ indicates higher magnification of the line wings than the peak, and  $RBI < 0$ indicates a higher magnification of the blue part of the line than the red part.}
\label{fig:indall}
\end{figure*}

\begin{figure*}[t]
  \resizebox{\hsize}{!}{\includegraphics*{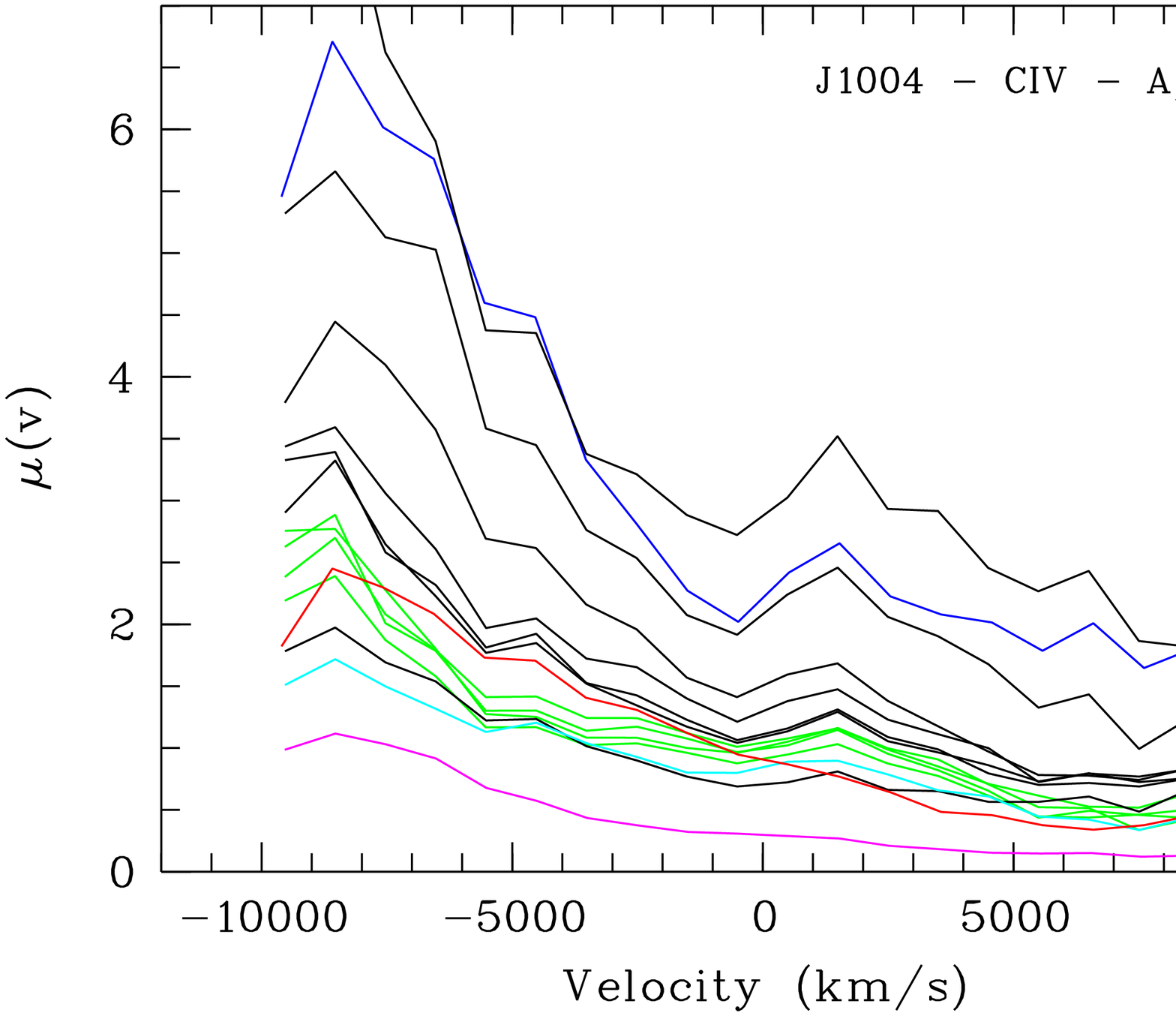}\includegraphics*{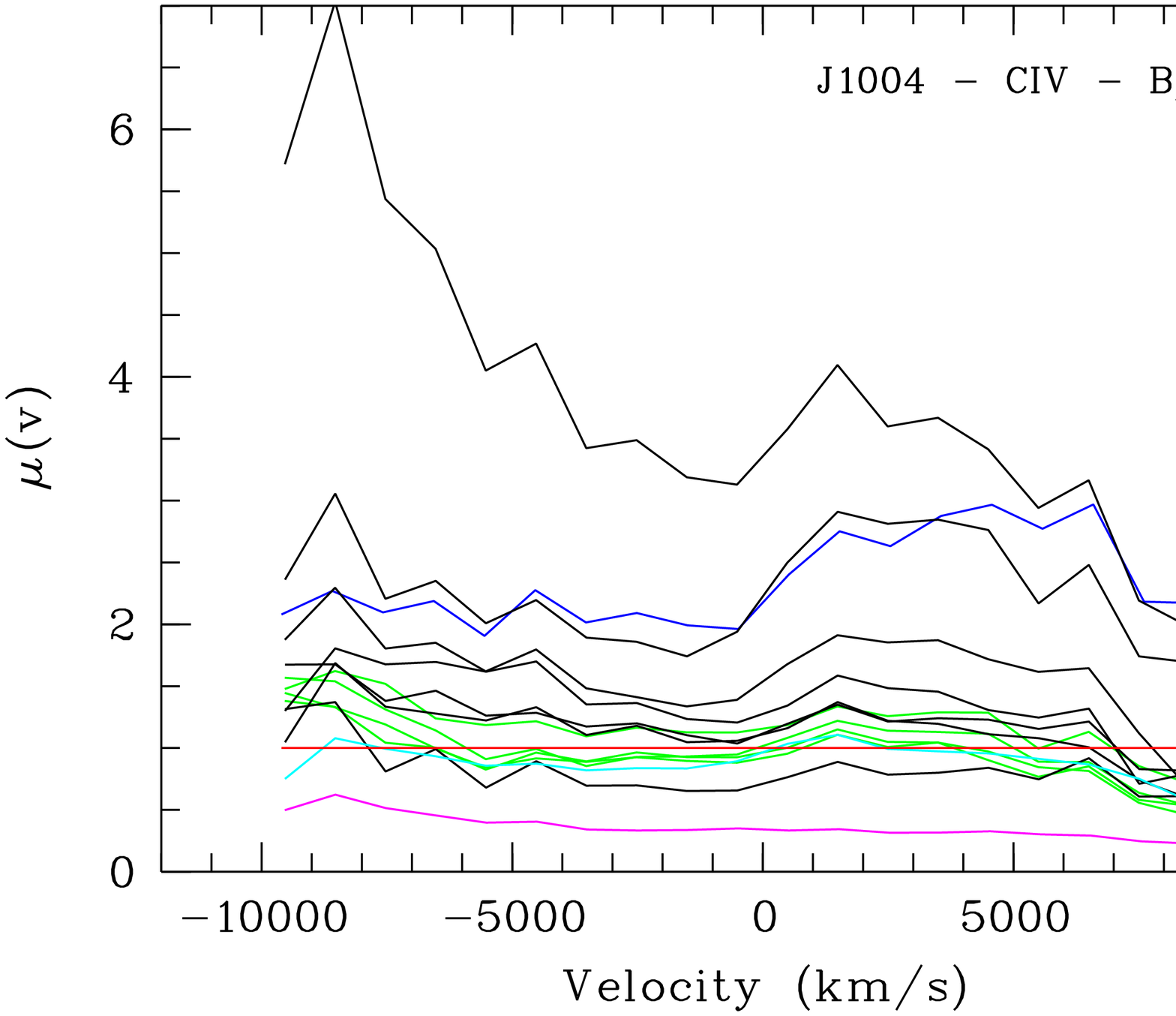}}\\
  \resizebox{\hsize}{!}{\includegraphics*{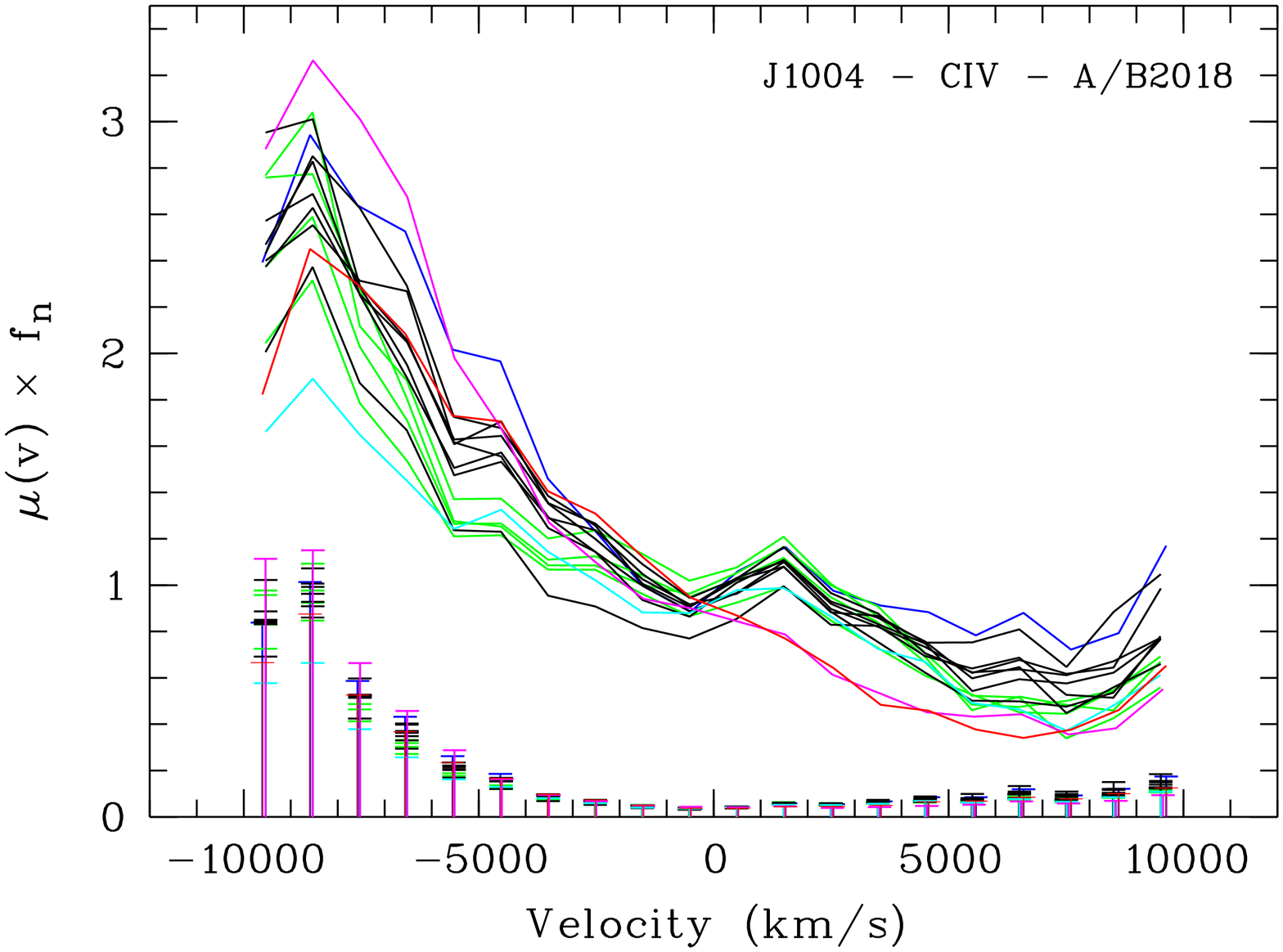}\includegraphics*{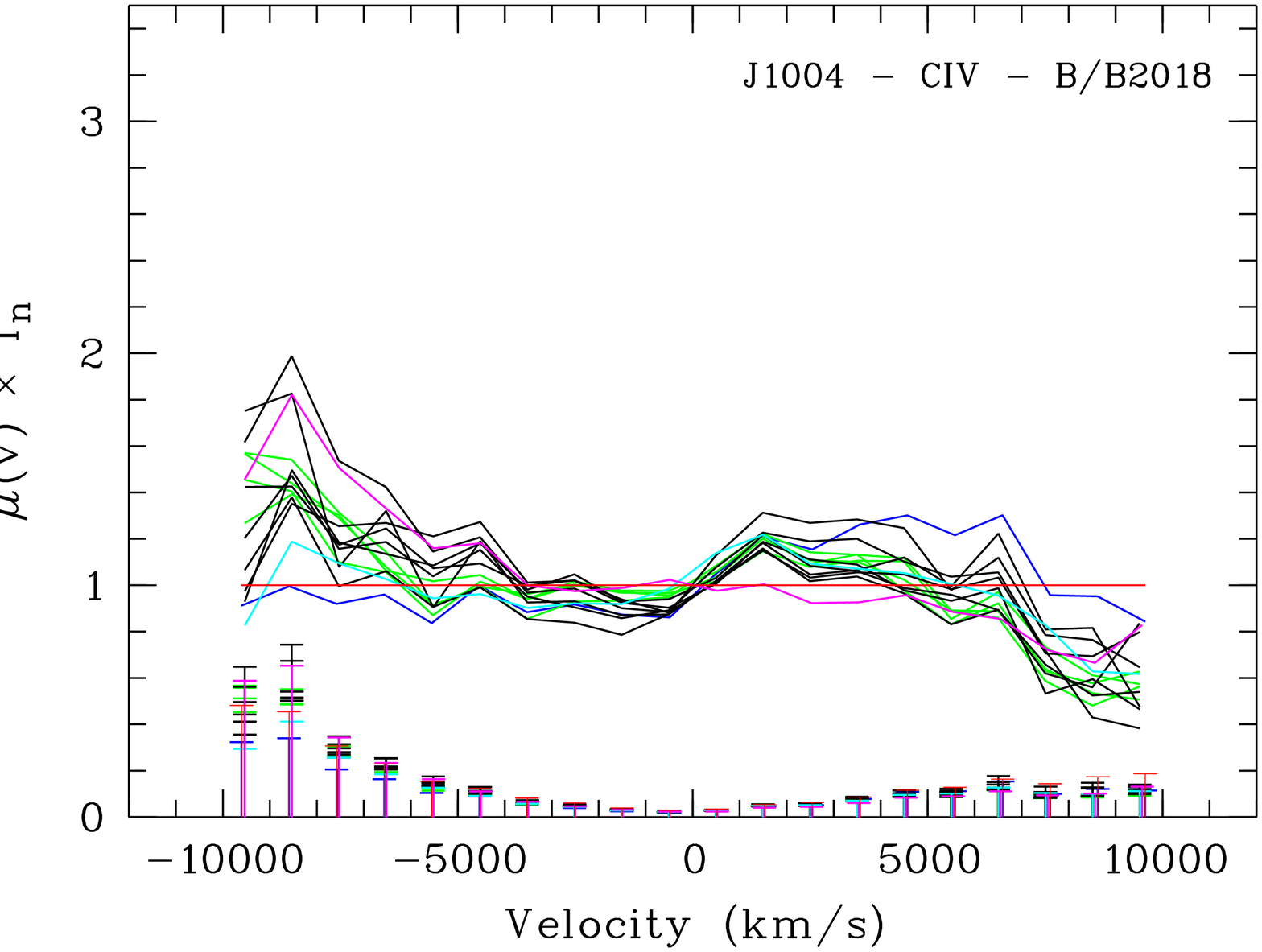}}\\
  \caption{Time series of $\mu(v)$  profiles of \ion{C}{iv} computed from the spectra of images A and B with respect to the 2018 spectrum of image B, unaffected by microlensing (top).  Same profiles are shown below, but normalized by the factor, $f_n$, defined in the text. The color-coding of the profiles and the error bars are the same as in Fig.~\ref{fig:muvall}. The A/B2018 $\mu(v)$ profiles are not affected by microlensing in B, while the B/B2018 $\mu(v)$ profiles are not affected by microlensing in A. The normalized A/B2018 $\mu(v)$ profiles have similar shapes at the different epochs.}
\label{fig:muvallr}
\end{figure*}

Figure~\ref{fig:profiles} shows the continuum-subtracted \ion{C}{iv} emission line profiles observed in 2003 and 2004 in images A and B. The continuum was measured in two windows on each side of the line profile, [3925,3980] and [4660,4715] \AA,\  which correspond to $[-22,-18]$ and $[+30,+34]$ 10$^3$ km~s$^{-1}$. It was then interpolated by a straight line under the line profile. The line profiles are superimposed according to different normalizations, either with respect to the line peak or to the blue wing. The flux enhancement in the blue wing of component A in May 2003, followed by its decrease later in 2003, and its recovery in March 2004 \citep{2004Richards,2004bRichards,2006Gomez} are best seen when normalizing the line profiles to the peak intensity. The enhancement and its variability were interpreted by a moving microlensing caustic \citep{2004Richards}, although the recurrence of the same feature has cast doubt on this interpretation \citep{2006Gomez}. Normalizing the line profiles to the blue wing tells another story: while the net enhancement of the blue wing with respect to the blue wing seen in image B is still present, the profile variations appear as global changes that include the line core. This led \citet{2006Lamer} to propose a scenario in which the blue wing is affected by a long-term, stable microlensing effect in image~A, and where the observed changes are due to intrinsic variability on a timescale of a few months. This scenario is supported by the fact that similar variations are seen in the image B spectrum (Fig.~\ref{fig:profiles}), as predicted by \citet{2006Lamer}, although it does not fully explain the extreme behavior of the May 2003 spectrum. These results illustrate the complex intertwining of microlensing and intrinsic variations. In the remainder of this section, we try to disentangle these two effects.

To characterize and quantify the line profile deformations induced by the microlensing effect, we considered the magnification profile $\mu(v)$, which is the macro-magnification-corrected ratio of the continuum-subtracted emission line flux densities observed in two different images. We also considered three indices:
1) $\mu^{BLR}$, integrated over the line profile; it essentially quantifies the total magnification of the line,
2) $WCI$, the wings-core index integrated over the $\mu(v)$ profile; when it is above or below a value of 1, it indicates whether the whole emission line is, on average, more or less affected by microlensing than its center, and 
3) $RBI$, the red-blue index integrated over the $\mu(v)$ profile; it takes non-null values when the effect of microlensing on the blue and red parts of the line is asymmetric.
A fourth index,  $\mu^{cont}$, measures the microlensing magnification of the continuum underlying the emission line. All these quantities were computed using the spectrum of one image affected by microlensing and the spectrum of a second image not or much less affected. When the spectra of both images are simultaneously recorded,  $\mu(v)$ and the indices are independent of quasar intrinsic variations that occur on timescales longer that the time delay between the two images. The indices $\mu^{cont}$ and $\mu^{BLR}$ are corrected for the image macro-magnification ratio; $RBI$ and $WCI$ are independent of this ratio. An exact definition of these quantities can be found in \citet{2017Braibant} or \citet{2021Hutsemekers}.

Since microlensing does not affect the radio emission, which originates from a more extended source than the optical emission,  in the following, we use the macro-magnification ratio, $M_{\rm A} / M_{\rm B} = 1.60 \pm 0.15,$ computed from the radio fluxes given in \citet{2021Hartley}. Figure~\ref{fig:exmuv} shows examples of $\mu(v)$ profiles computed at four epochs from images A and B. As a flux ratio, $\mu(v)$ can be extremely noisy in the wings of the emission lines where the flux density reaches zero, so that it is necessary to cut the faintest parts of the line wings.  Thus, we only considered the parts of the line profiles whose flux density is above $l_{\rm cut} \times F_{\rm peak}$, where $F_{\rm peak}$ is the maximum flux in the line profile and $l_{\rm cut}$ is fixed around 0.1. In practice, this corresponds to cut the profiles between $-10^4$ and $+10^4$ km s$^{-1}$ at all epochs. To increase the signal-to-noise ratio, we binned $\mu(v)$ into 20 spectral elements, which also corresponds to the spectral resolution of the line profiles produced by the microlensing simulations (see Sect.~\ref{sec:modeling}).

A proper interpretation of the microlensing effect in one image requires a reference image that is not affected by microlensing, as well as a short time delay between the two images, so as to rule out the possibility that observed spectral differences could be attributed to a delayed, short timescale intrinsic variability. The time delay between images B and A is 40.6 $\pm$ 1.8 days \citep{2008Fohlmeister}, which is small with respect to the timescale of the intrinsic variations shown in \citet{2008Fohlmeister} and \citet{2022Munoz}. We may therefore assume that there is only a negligible delay between images A and B when a significant intrinsic variation occurs; however, we ought to keep in mind that for some rare events, this might not be the case. This assumption is supported by simulations that show that (statistically) intrinsic variability is unlikely to mimic microlensing-induced line deformations if the time delay is shorter than about 50 days \citep{2012Sluse}.  

In 2018 (and only in 2018), the \ion{C}{iv} continuum and line profile of images B and C are nearly ideally superimposed with a constant factor, indicating the absence of microlensing in both images B and C (Fig.~\ref{fig:muvbc}).  The constant factor $M_{\rm B} / M_{\rm C} = 1.40 \pm 0.14$ is in agreement, within the uncertainties, with the macro-magnification ratio derived from radio observations, $M_{\rm B} / M_{\rm C} = 1.10 \pm 0.13$ \citep{2021Hartley}, keeping in mind that the optical ratio could be slightly off because the A-B and C-D spectra were not obtained simultaneously, making the relative flux calibration of B and C more uncertain \citep{2020Popovic}. The B/C magnification profile $\mu(v)$ is flat (Fig.~\ref{fig:muvbc}). The microlensing indices $\mu^{cont}$,  $\mu^{BLR}$, and $WCI$ measured between images B and C are around one, and $RBI$ around zero, consistently indicating the absence of any differential microlensing affecting the \ion{C}{iv} broad emission line in images B and C. Without microlensing, the line profiles appear quasi-symmetric. 

By comparing the spectra of images A and B at that epoch using B as the non-microlensed reference, we can thus unambiguously characterize the microlensing effect observed in image A. We find that the continuum of image A is significantly demagnified with $\mu^{cont} = 0.59 \pm 0.10$, differential dust extinction being negligible \citep{2012Motta}. The \ion{C}{iv} emission line profile is strongly distorted (Fig.~\ref{fig:exmuv}): the blue part of the profile is magnified by up to a factor of two, the red part demagnified by almost the same factor, reaching the continuum demagnification  $\mu^{cont}$ at the highest velocities, while the line core is essentially unaffected by the microlensing effect. 

As seen in Fig.~\ref{fig:exmuv} and \ref{fig:muvall}, the $\mu(v)$ profiles derived from images A and B retain roughly the same shape at the different epochs, with significant variability in the blue and red parts of the profile. The fact that $\mu(v=0)$ remains stable at all epochs confirms that these variations are not due to delayed, short-timescale intrinsic variations that would otherwise mimic the microlensing magnification pattern. Since $\mu(v)$ is independent of long-term intrinsic profile variations, the $\mu(v)$ profile variability is thus likely due to microlensing variations in either image A or image B, or both. Time series of the four indices are shown in Fig.~\ref{fig:indall}. Consistently, at all epochs, $WCI > 1$ indicates higher magnification of the line wings than the peak, and  $RBI < 0$ indicates higher magnification of the blue part of the line than the red part. $\mu^{BLR}$ remains nearly constant, namely, around one, indicating no significant magnification of the line profile as a whole, the demagnification of the red part of the line profile counterbalancing the magnification of the blue part. $\mu^{cont}$ shows significant variation, reaching $\mu^{cont} \simeq $ 0.6 - 0.7 after 2006. As for the $\mu(v)$ profile, the variability of the indices is likely due to microlensing variability in component A, B, or both.

In Fig.~\ref{fig:muvallr}, we show $\mu(v)$ computed from the spectra of image A at all epochs with respect to the spectrum of image B obtained in 2018 (hereafter, B2018), which is unaffected by microlensing. We also compute $\mu(v)$ from the spectra of image B at all epochs with respect to its 2018 spectrum. As expected, variations of the B/B2018 $\mu(v)$ profile are observed. They are due to intrinsic variations between 2003 and 2018, microlensing in image B, and absolute flux calibration errors. We then normalize the $\mu(v)$ profiles by the factor $f_n = F_{\rm BEL}({\rm B2018}) / F_{\rm BEL}({\rm B}),$ where $F_{\rm BEL}$ is the flux integrated over the continuum-subtracted \ion{C}{iv} line profile. This normalization corrects for calibration errors as well as global intrinsic and microlensing variations, but not for line profile deformations. After normalization, time-dependent $\mu(v)$ profile deformations are still present in image B, due to intrinsic or microlensing velocity-dependent variations in image B. Similarly, time-dependent $\mu(v)$ profile deformations are present in image A with respect to B2018; in this case, they are due to intrinsic or velocity-dependent microlensing variations in image~A.

If only of intrinsic origin, these variations would disappear from the A/B $\mu(v)$ profiles derived from images A and B observed at the same epoch (Fig.~\ref{fig:muvall}), resulting in a single A/B $\mu(v)$ profile for all epochs. Since this is not observed, microlensing variations are thus present in both images A and B, affecting the A/B2108 and B/B2018 $\mu(v)$ profiles separately and contaminating the A/B $\mu(v)$ profiles seen in Fig.~\ref{fig:muvall}. The normalized A/B2018 $\mu(v)$ profiles, which are not affected by microlensing in B, show less scatter than the A/B $\mu(v)$ profiles as well as more similar shapes through the different epochs, suggesting that they are closer to the true microlensing effect that occurs in image A.

In summary, we can see that although they are variable, the microlensing-induced distortions of the \ion{C}{iv} line profile seen in image A appear remarkably similar over a period of 15 years. The distortions are characterized by a magnification of the blue part of the profile and a demagnification of the red part, with only a small demagnification of the line core. The magnification $<\! \mu' \!\!>$ = $<\! \mu\times f_n \!\!>$  averaged over the different epochs (Fig.~\ref{fig:muvallr}) approximately decreases as $\log_{10} <\! \mu' \!\!>$ $ \simeq -0.4 \times (v / 10000$ km~s$^{-1}$). The stronger demagnification observed in the velocity range 1000-5000 km~s$^{-1}$ in the two spectra obtained after 2008 (in magenta and red) may indicate a long-term change (see also $RBI$ in Fig.~\ref{fig:indall}). Microlensing is also at work in image B, albeit more weakly.
In the following, we compare the $\mu(v)$ profile and the indices to simulations, focusing on the 2018 data for which the microlensing effect can be unambiguously characterized.

\section{Microlensing simulations}
\label{sec:modeling}

\subsection{Methodology and model parameters}

We computed the effect of gravitational microlensing on the broad emission line profiles by convolving in the source plane the emission from representative BLR models with microlensing magnification maps. For the BLR models, we considered a rotating Keplerian disk (KD), as well as biconical polar (PW) and equatorial (EW) radially accelerated winds characterized by inclinations with respect to the line of sight of $i$ = 22\degr, 34\degr, 44\degr, 62\degr.  We considered $11$ BLR inner radius values: $r_{\text{in}}$ = 0.025, 0.05, 0.075, 0.1, 0.125, 0.15, 0.175, 0.2, 0.25, 0.35, and 0.5 $r_E$, where $r_E$ is the microlensing Einstein radius in the source plane. The outer radius is fixed to $r_{\text{out}} = 10 \, r_{\text{in}}$.  The BLR models are assumed to have an emissivity $\epsilon = \epsilon_0 \, (r_{\text{in}}/r)^q$ that either sharply decreases with increasing radius, $q=3$, or more slowly, $q=1.5$. Twenty BLR monochromatic images forming a data cube and corresponding to twenty spectral bins in the line profile are produced, using the radiative transfer code STOKES \citep{2007Goosmann,2012Marin,2014Goosmann}. Each model also contains a continuum-emitting uniform disk seen under the same inclination as the BLR, and with an outer radius fixed at $11$ different values: $r_s =$ 0.02, 0.03, 0.05, 0.07, 0.1, 0.15, 0.2, 0.25, 0.3, 0.4, and 0.5 $r_E$.  Only models with $r_{\text{in}} \geq r_s$ are considered.  For a full description of the models and additional details, we refer to \citet{2017Braibant}.

Modeling the effect of microlensing on the BLR is achieved using a magnification map specific to the image A of J1004+4112.  The magnification map is computed using the \texttt{microlens} ray-tracing code \citep{1999Wambsganss}, considering a total convergence of $\kappa = $ 0.763  with a shear $\gamma =$ 0.3. The fraction of matter in compact objects is  arbitrarily taken to be lower than usual because the system is lensed by a cluster: $\kappa_{\star} / \kappa $ = 3\%. These values are from \citet{2017Guerras} based on the model of \citet{2010Oguri} and they are in agreement with more recent modeling by \citet{2022Fores}. Since $\kappa_{\star} / \kappa $ is poorly known, we tested the robustness of the results using $\kappa_{\star} / \kappa $ = 1\% and $\kappa_{\star} / \kappa $ = 7\%, the latter value being more typical of the fraction of compact objects in lens galaxies \citep{2009Mediavilla}. The map extends over a $200 \, r_E \times 200 \, r_E$ area of the source plane and is sampled by $20000 \times 20000$ pixels. To investigate the impact of preferential alignment between the symmetry axis of the BLR models and the caustic network, the maps are rotated clockwise by $\theta$ = 15\degr, 30\degr, 45\degr, 60\degr, 75\degr,\ and 90\degr\ with respect to the BLR model axis. Here, $\theta$ = 0\degr\ corresponds to the caustic elongation and shear direction perpendicular to the BLR model axis. After rotation, only the central $10000 \times 10000$ pixel part of the map is used.

Distorted line profiles are obtained from the convolution of the monochromatic images of the BLR by the magnification maps and computed for each position of the BLR on the magnification maps, generating $\sim 10^8$ simulated profiles per map and BLR model. In previous works, we only considered the indices $\mu^{cont}$, $\mu^{BLR}$, $RBI$, and $WCI$ for the comparison with the observations. In the present work, we modified the code to accelerate the convolution and save disk space so that we can also use the full $\mu(v)$ magnification profiles with 20 spectral elements. For a proper comparison, the wings of the simulated line profiles are truncated as the observed ones, so that only the parts of the profiles with a flux density larger than $0.1 \times F_{\rm peak}$ are considered. After truncation, the simulated line profiles, and thus the data cubes, are finally rescaled to match the observed $\mu(v)$ profile in velocity. Simulations show that the final probabilities of the different BLR models are robust with respect to small changes of the adopted flux threshold. 

\subsection{Probability of the different BLR models}
\label{sec:proba}

\begin{figure*}[t]
\resizebox{\hsize}{!}{\includegraphics*{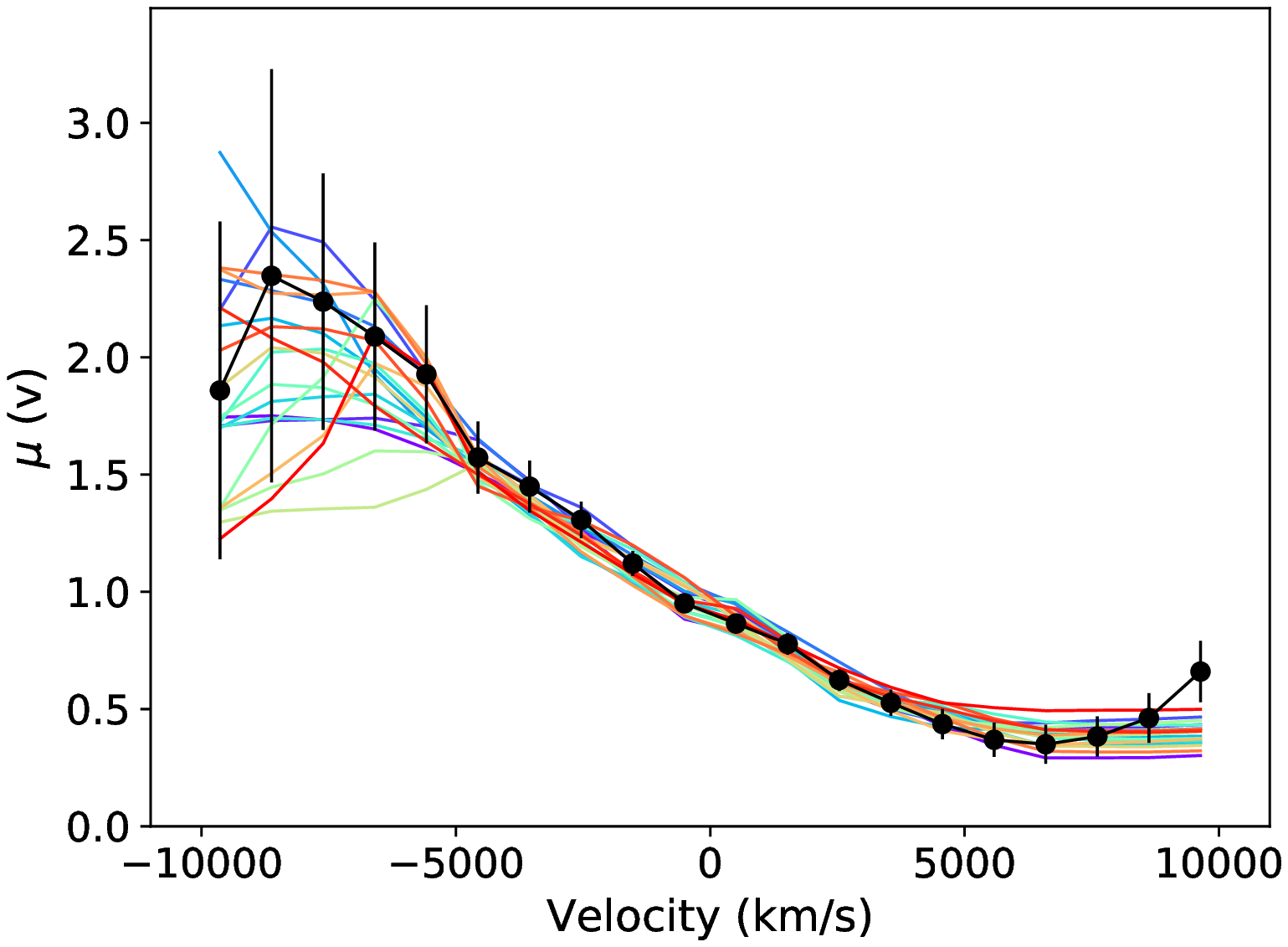}\includegraphics*{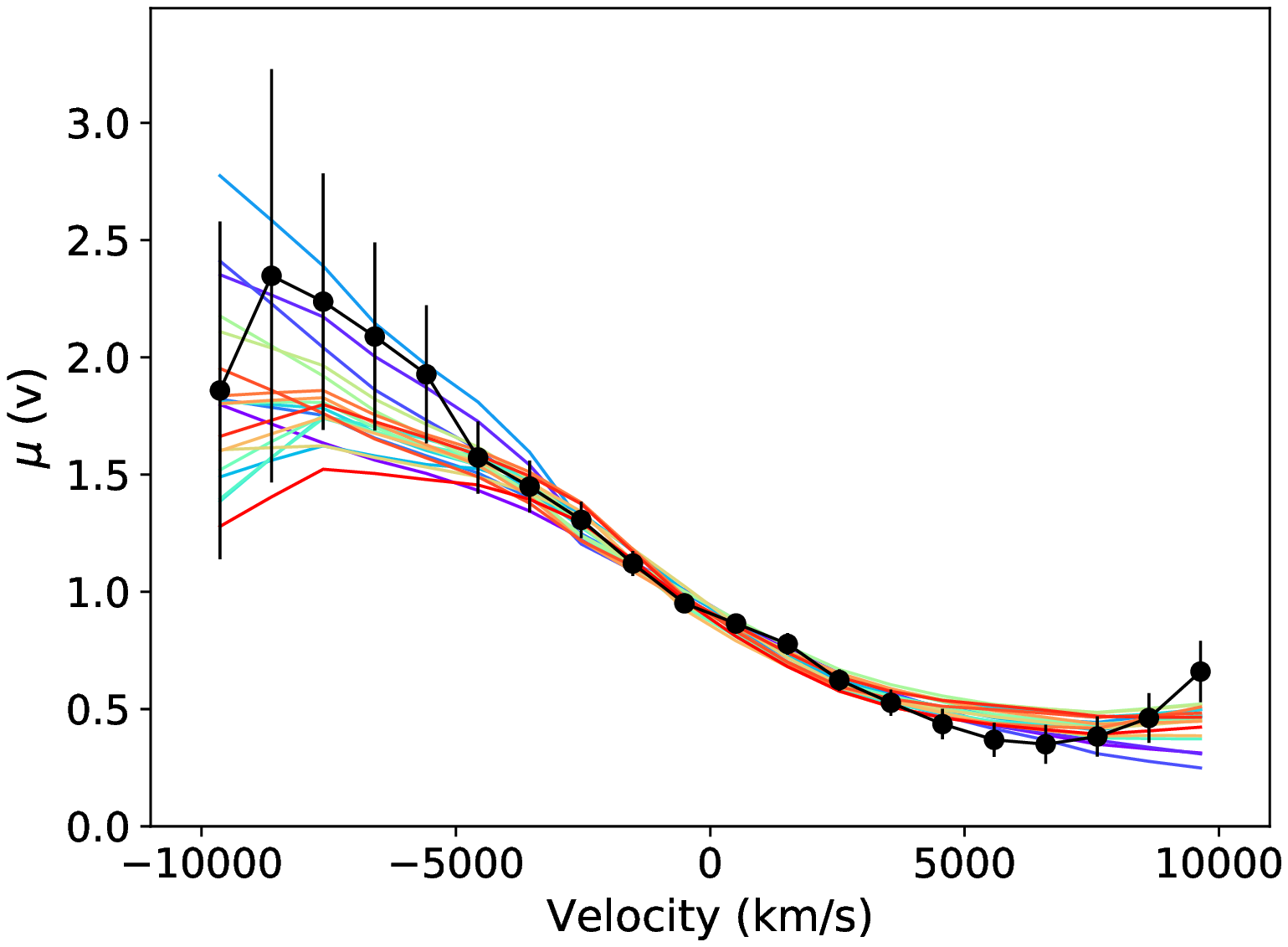}}\\
\caption{Examples of 20 simulated $\mu(v)$ profiles (in color) that fit, with $\rchi^{2} / n_{\rm d.o.f.} \leq 1.2$, the $\mu(v)$ magnification profile measured in 2018 for the \ion{C}{iv} emission line (in black). The simulated profiles are computed with $i = 34\degr$, $q = 3$, $r_{\rm in} = 0.15 \, r_{\rm E}$, and $r_s = 0.03 \, r_{\rm E}$. The left and right panels show profiles from the KD and EW models computed with the magnification map oriented at  $\theta$ = 60\degr\ and $\theta$ = 30\degr,  respectively.}
\label{fig:fitmuv}
\end{figure*}

In order to identify the BLR models that provide the best fit to the observations, we computed the relative posterior probability that a given model $(G, i)$ (where $G$ = KD, PW, or EW, and $i$ = 22\degr, 34\degr, 44\degr, or 62\degr) can reproduce the four observables $\mu^{cont}$, $\mu^{BLR}$, $RBI$, and $WCI$. The method is  described in detail in \citet{2019Hutsemekers}. In practice, we computed the likelihood of the observables for each set of parameters characterizing the simulations. We then marginalized the likelihood over $r_s$, $r_{\text{in}}$, and $q$, as well as over the microlensing parameters. Since the different BLR models share the same parameters and associated priors, we can quantify their relative efficiency to reproduce the data by comparing their likelihoods, that is, by normalizing the marginalized likelihood by the sum of the likelihoods associated to each model, $G,$ for each inclination, $i$. This procedure yields the relative probability of the different models.

The probabilities of the different models are similarly computed using $\mu^{cont}$ and the 20 spectral elements of $\mu(v)$ instead of the three indices used to characterize the line profile magnification. However, given the higher number of degrees of freedom and the non-continuous nature of the model parameters, the previously used likelihood estimator $L_G = \exp \, (-0.5 \,\, \rchi^{2})$, which is very sensitive to small, rare $\rchi^{2}$ values, may provide unreliable results. In particular, observed $\mu(v)$ profiles comparable within the uncertainties can result in very different probabilities of the models. To circumvent this problem, we considered the likelihood estimator used by \citet{2004Kochanek}:
\begin{equation}
  L_K \, = \, \Gamma \, [\frac{n_{\rm d.o.f.}-2}{2}, \frac{\rchi^2}{2\,f^2_0}] \, ,
\end{equation}
where $n_{\rm d.o.f.} = 19$ is the number of degrees of freedom, and $f_0$ a free parameter. This likelihood estimator is essentially flat for small $\rchi^{2} / n_{\rm d.o.f.}$ values, and behaves like the $L_G$  estimator for higher \ $\rchi^{2}$ values. With the $L_K$ estimator and $f_0$ fixed to one, the resulting model probabilities were found to be stable with respect to small changes of the observed $\mu(v)$ profile.

\section{Results}
\label{sec:results}

\begin{figure*}[h]
\resizebox{\hsize}{!}{\includegraphics*{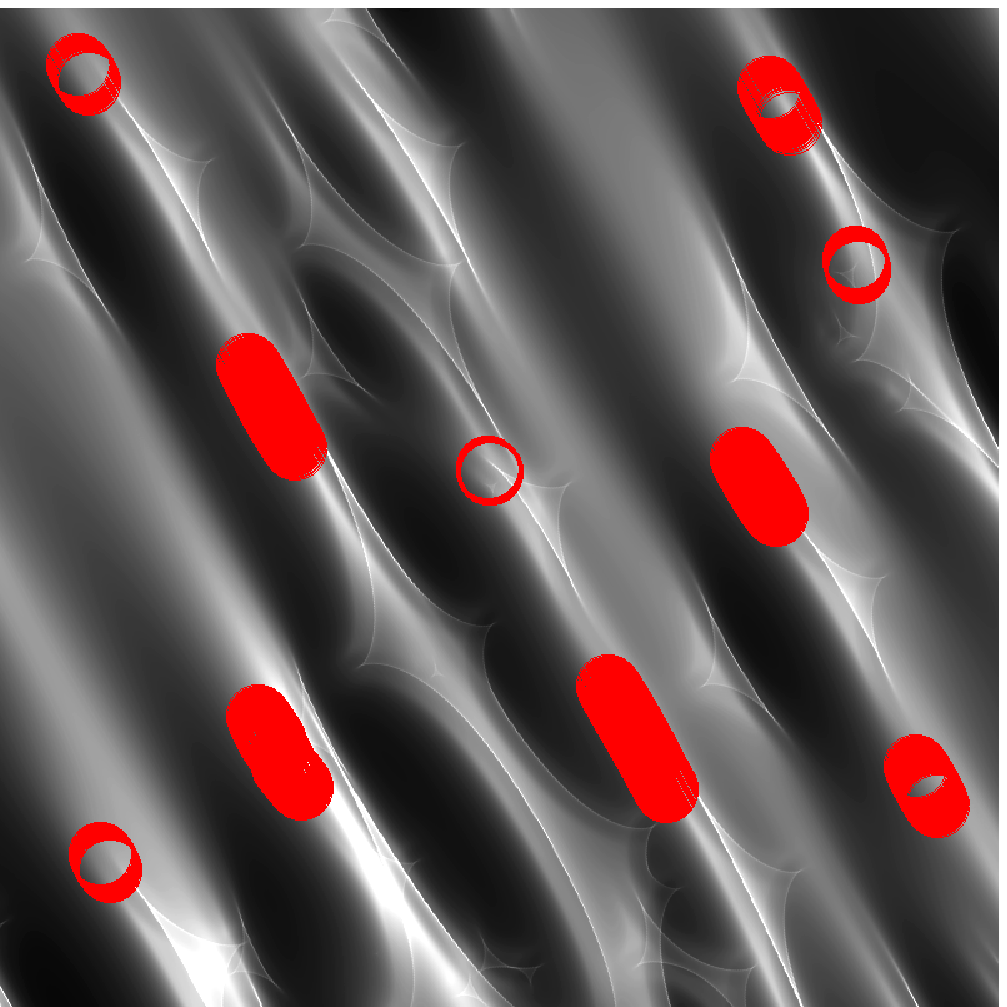}\ \includegraphics*{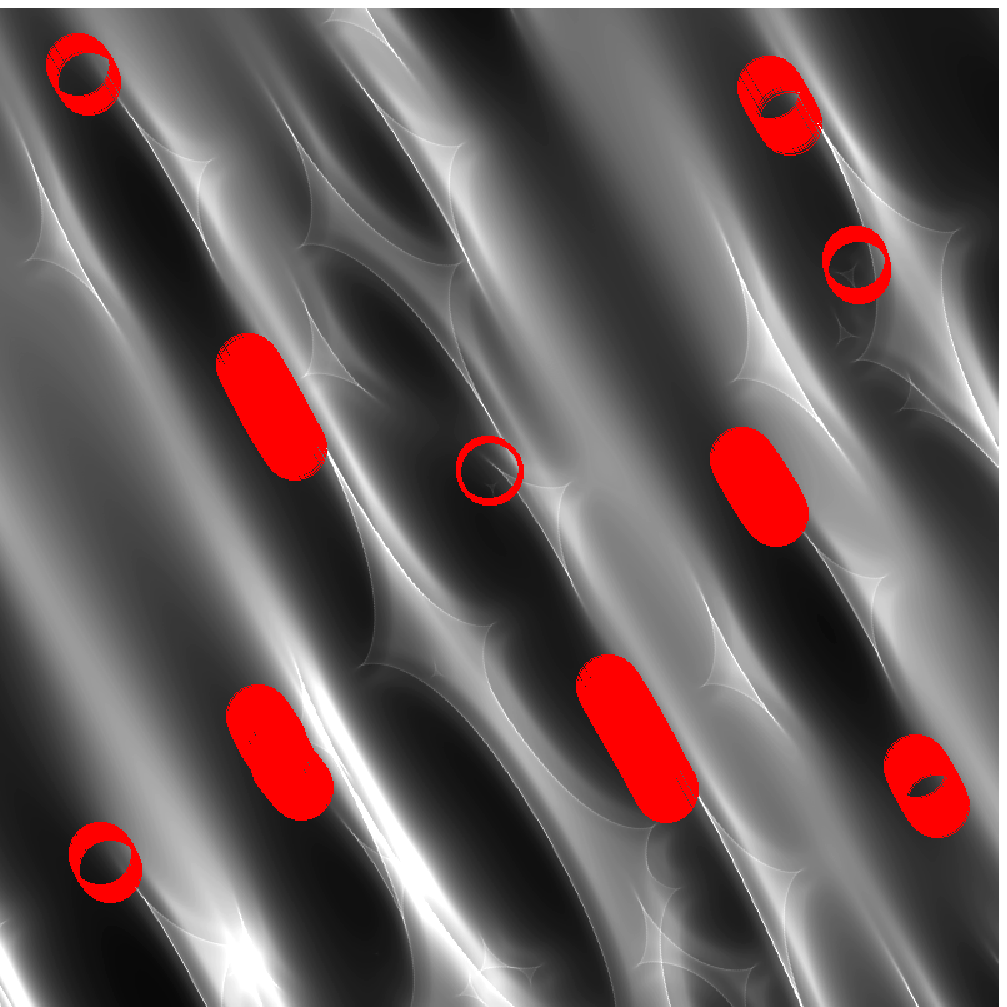}}
\caption{Examples of positions on the magnification map where the computed line profile deformations fit the observed one with $\rchi^{2} / n_{\rm d.o.f.} \leq 1.2$ and where the computed indices match the observed ones within the error bars. The exact positions are at the center of the circles, the radius of which represents the BLR half-light radius $r_{1/2} = 1.64 \, r_{\rm in} = 0.25 \, r_{\rm E}$. The left panel shows the map convolved by the part of the BLR at the origin of the blue part of the line profile, while the right panel shows the map convolved by the part of the BLR at the origin of the red part of the line profile. The unconvolved magnification map is superimposed and seen as narrow caustics. The considered model is KD with $i = 34\degr$, $q = 3$, $r_{\rm in} = 0.15 \, r_{\rm E}$, and $r_s = 0.03 \, r_{\rm E}$. The magnification map, computed with $\kappa_{\star} / \kappa $ = 3\%, is oriented at  $\theta$ = 60\degr. The size of the illustrated part of the map is $8 \, r_{\rm E} \times 8 \, r_{\rm E}$. Similar patterns are observed throughout the full map, as well as for the EW model with $\theta \leq$  30\degr. }
\label{fig:mappos}
\end{figure*}

For the comparison with the microlensing simulations, we used the $\mu(v)$ magnification profile measured from the spectra of images A and B obtained in 2018, with image B being unaffected by microlensing at that epoch. The indices measured from this data set are $\mu^{cont} = 0.59 \pm 0.10$, $\mu^{BLR} = 1.01 \pm 0.08$, $WCI = 1.24 \pm 0.11$, and $RBI = -0.50 \pm 0.03$.

In Fig.~\ref{fig:fitmuv}, we show examples of simulated $\mu(v)$ profile for the KD and EW models (the most likely BLR models; see  Table~\ref{tab:proba}). As can be seen, the observed $\mu(v)$ profile is reproduced by a number of simulated profiles, without the need for more sophisticated BLR models, in particular, models involving two BLRs of different sizes. Figure~\ref{fig:mappos} shows typical locations of the BLR on the magnification map, at which the observed line profile deformations and the continuum demagnification are reproduced. As anticipated, one of the high-velocity parts of the BLR coincides with high-magnification caustics, with the other part and the continuum source lying in a demagnification region. The microlensing effect appears dominated by the distance of the different parts of the BLR with respect to the caustics rather than by the velocity-dependent BLR size. Indeed, the effective radius of the BLR is smaller at high velocities for the KD model, while it is larger for the EW model (at least for inclinations lower than about 60\degr; \citealt{2017Braibant}). If the microlensing magnification were dominated by the source size, we would expect higher magnification in the line wings for the KD models and in the line core for the EW models. On the contrary, both models reproduce the observed $\mu(v)$ profile that shows high (de)magnification in the wings only. Moreover, the change of the effective BLR radius with the velocity remains smaller than a factor three \citep{2017Braibant}. Since the magnification varies as $1/ \! \sqrt{r}$ when close to a caustic,  with $r$ denoting the source radius \citep{1992Schneider}, the change in radius would induce at most a factor of$\sqrt{3}$ in magnification, which is lower than the magnification ratio observed between the different parts of the line profile.

Similar microlensing effects are obtained from close positions on the maps, which may extend over regions larger than one $r_{\rm E}$ (Fig.~\ref{fig:mappos}). The timescale to cross one $r_{\rm E}$ is about 10 years \citep{2011Mosquera}, so that the observation of similar line-profile deformations over 15 years is plausible. Since crossing caustics would typically produce blue-red sequences of magnification in the line profiles \citep{2007Abajas,2017Braibant}, the similarity of the magnification profiles indicates the absence of caustic crossing during this time interval. Interestingly, tracks that do not cross caustics are more likely to occur in lens systems with stretched caustic networks like J1004+4112 (Fig.~\ref{fig:mappos}) than in systems with more randomly oriented caustics such as Q2237+0305 (see \citealt{2021Hutsemekers}).

\subsection{Geometry and kinematics of the BLR}
\label{sec:geometry}

The probabilities of the different models of the \ion{C}{iv} BLR are given in Table~\ref{tab:proba}.  We found no significant differences using the magnification maps computed with the different $\kappa_{\star} / \kappa $ values. The results obtained using the full $\mu(v)$ profile are in good agreement with those derived from the integrated indices.  The model selection is mostly driven by the large $RBI$ value, as seen in $WCI-RBI$ diagrams \citep{2017Braibant,2019Hutsemekers,2021Hutsemekers}. Since the probablities were found to strongly depend on the orientation of the magnification maps with respect to the BLR axis, we computed them individually for the map orientations $\theta \leq 30\degr$ and $\theta \geq 60\degr$. In the first case, EW models are strongly favored, while in the second case, KD models are the only possibility. In all cases, the PW models, within the framework of which high RBI values are more difficult to reproduce \citep{2017Braibant,2019Hutsemekers,2021Hutsemekers}, are much less likely and can thus be rejected. The dichotomy in the probabilities of the KD and EW models with respect to the magnification map orientation is due to the fact that the caustic network is especially stretched in one direction (Fig.~\ref{fig:mappos}); in addition, in the plane of the sky, the high-velocity regions of the BLR are perpendicular to the BLR axis in the KD models, whereas they are parallel in the EW models \citep{2017Braibant}. Unfortunately, no constraint on the radio jet orientation can be derived from the Very Large Array maps shown in \citet{2021Hartley}. Using the jet position angle as a proxy of the BLR axis orientation, it would have been possible to disentangle the KD and EW models.

\begin{table}[t]
\caption{Probability (in \%) of the \ion{C}{iv} BLR models}
\label{tab:proba}
\centering
\begin{tabular}{lccccccc}
\hline\hline
     & \multicolumn{3}{c}{$\theta \leq 30\degr$} &  & \multicolumn{3}{c}{$\theta \geq 60\degr$} \\
\hline
     \multicolumn{8}{c}{Using the 4 indices} \\
\hline
          & KD & PW & EW &  & KD & PW & EW  \\
\hline
          22\degr         &  1 &  0 & 12    & & 20 &  0 &  0  \\
          34\degr         &  1 &  0 & 18    & & 24 &  0 &  0  \\
          44\degr         &  2 &  0 & 20    & & 26 &  0 &  0  \\
          62\degr         &  6 & 17 & 23    & & 30 &  0 &  0  \\
          All $i$         & 10 & 17 & 74    & & 100 & 0 &  0    \\
\hline
     \multicolumn{8}{c}{Using $\mu^{cont}$ and the $\mu(v)$ profile} \\
\hline
          & KD & PW & EW &  & KD & PW & EW  \\
\hline
          22\degr         & 0 &  0 &  17    & & 51 &  0 &  0  \\
          34\degr         & 0 &  0 &  27    & & 31 &  0 &  0  \\
          44\degr         & 0 &  0 &  24    & & 16 &  0 &  0  \\
          62\degr         & 0 &  2 &  29    & &  2 &  0 &  0  \\
          All $i$         & 0 &  2 &  97    & & 100 & 0 &  0     \\
\hline
\end{tabular}
\end{table}

\subsection{Size of the BLR}
\label{sec:size}

\begin{figure}[t]
\resizebox{\hsize}{!}{\includegraphics*{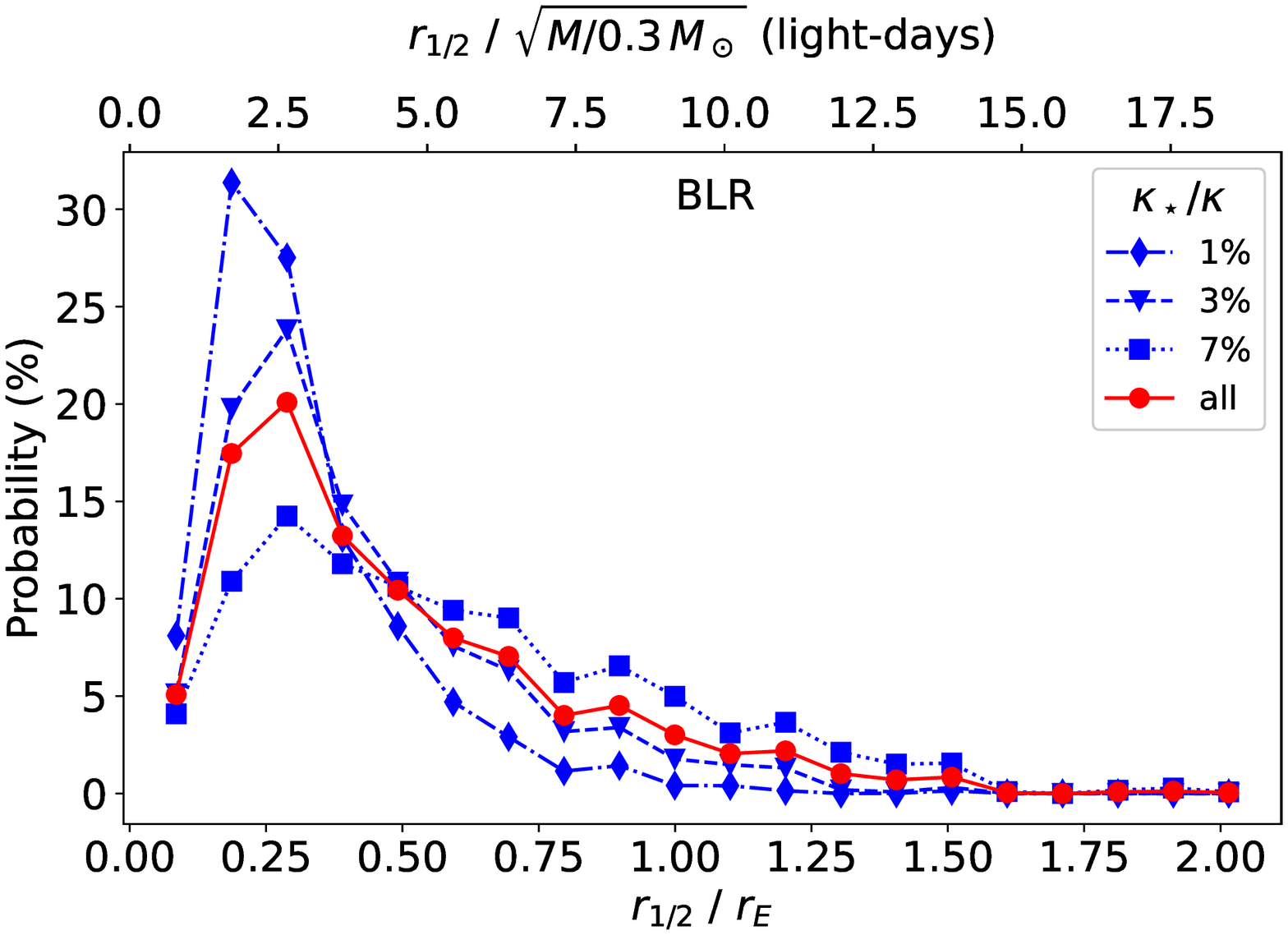}}
\caption{Relative probabilities (in percent) of the BLR half-light radius $r_{1/2}$ in $r_E$ and light-day units, considering the magnification maps computed with $\kappa_{\star} / \kappa$ = 1, 3, and 7\% (in blue). The probability distribution obtained by marginalizing over all $\kappa_{\star} / \kappa$ is shown in red.}
\label{fig:rbhl}
$ $ \\
\resizebox{\hsize}{!}{\includegraphics*{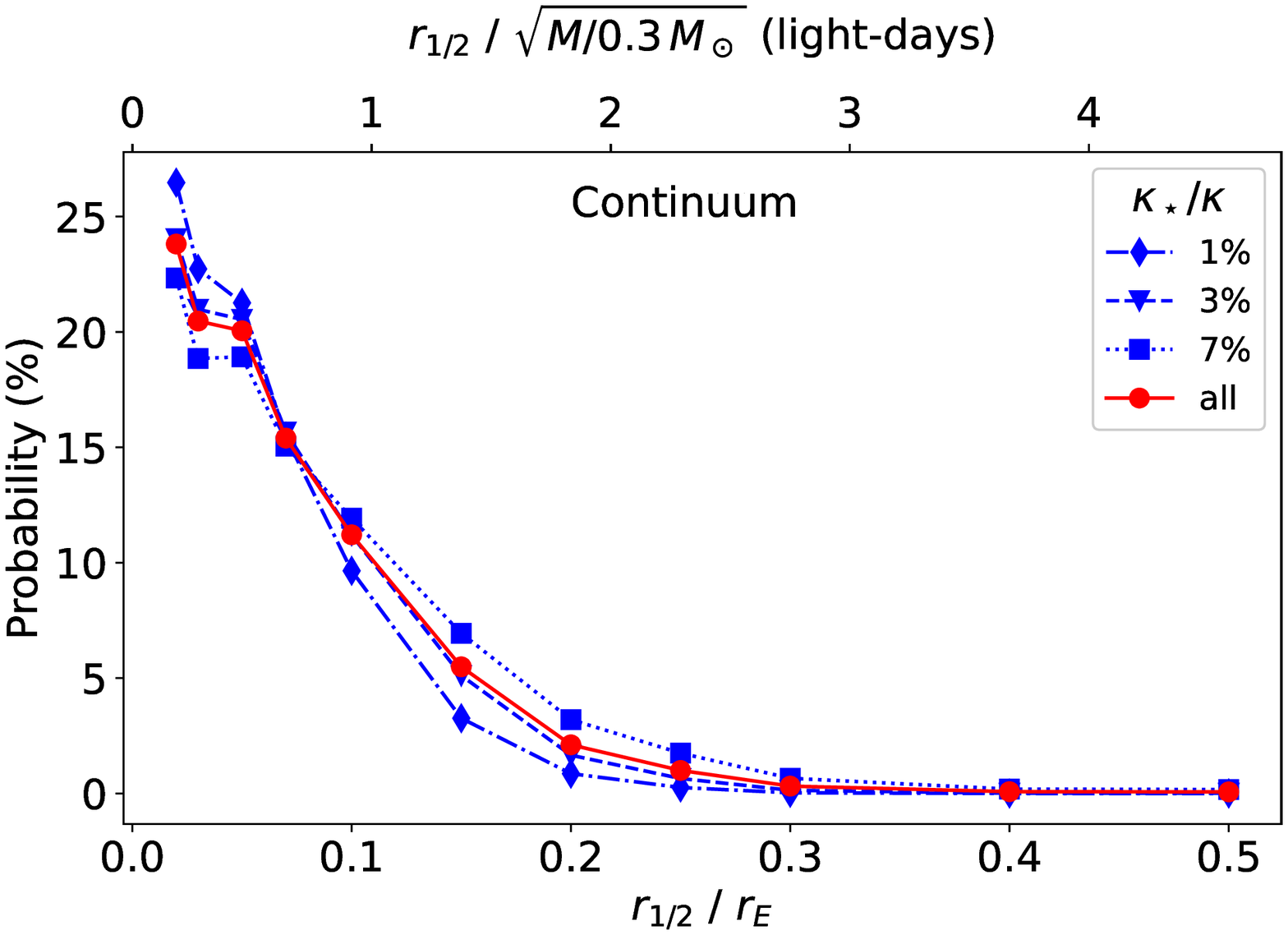}}
\caption{Same as Fig.~\ref{fig:rbhl} but for the continuum half-light radius.}
\label{fig:rchl}
\end{figure}

By marginalizing over all parameters but $r_{\text{in}}$, we can estimate the most likely BLR radius. Since $r_{\text{in}}$ does not properly represent the size of the BLR, which also depends on the light distribution, we computed the half-light radii for the different models as in \citet{2021Hutsemekers}. The resulting probabilities of the half-light radii, $r_{1/2}$, are shown in Fig.~\ref{fig:rbhl}. The probability distributions depend on $\kappa_{\star} / \kappa$, but the most likely value appears to be independent of it.  This value is equal to $r_{1/2} = 0.3^{+0.22}_{-0.18} \, r_E$, with the uncertainties corresponding to a 68\% confidence interval estimated for $\kappa_{\star} / \kappa$ = 3\%\footnote{We also computed the flux-weighted mean radius which is identical to the half-light radius within the uncertainties, as in the case of Q2237+0305 \citep{2021Hutsemekers}.}.  Similar values are obtained when  marginalizing over all values of $\kappa_{\star} / \kappa$. To compute the Einstein radius, we adopted a flat $\Lambda$CDM cosmology with $H_0 = 68$ km~s$^{-1}$ Mpc$^{-1}$ and $\Omega_m$ = 0.31. The source and lens redshifts are $z_S$ = 1.734 and $z_L$ = 0.68, respectively \citep{2003Inada}, so that the Einstein radius is $r_E$ = 9.17 $\sqrt{ M / 0.3 M_{\odot}}$ light-days in the source plane. For an average microlens mass of 0.3 M$_{\odot}$, we obtained:   $r_{1/2} = 2.8^{+2.0}_{-1.7}$ light-days. This value is unchanged when considering the map orientations $\theta \leq 30\degr$ or $\theta \geq 60\degr$ separately or when considering specific model inclinations. Similarly, we can estimate the half-light radius of the continuum source using  $r_{1/2} = r_s / \sqrt{2} $ for a uniform disk. The probability distribution is shown in Fig.~\ref{fig:rchl}, only providing an upper limit,  $r_{1/2} < 1.8$ light-days (95\% confidence level) at $\lambda_{\rm rest} = 1550 \AA$, consistently smaller than the BLR size.  This upper limit translates to $r_{1/2} < 2.9$ light-days at 2400 $\AA$, assuming  $r_{1/2} \propto \lambda ^{p} $ with $p = 1.1$ \citep{2012Motta}. It is in agreement with the continuum source size estimated by  \citet{2008Fohlmeister} using the light curve fitting method, $r_{1/2} = 0.8^{+0.8}_{-0.4}$ light-days at $\lambda_{\rm rest} = 2300 \AA$, but smaller than the size computed by \citet{2016Fian} based on the single epoch method, $r_{1/2} = 4.2^{+3.2}_{-2.2}$ -- $8.7^{+18.5}_{-5.5}$ light-days at $\lambda_{\rm rest} = 2407 \AA$.

\begin{figure}[t]
\resizebox{\hsize}{!}{\includegraphics*{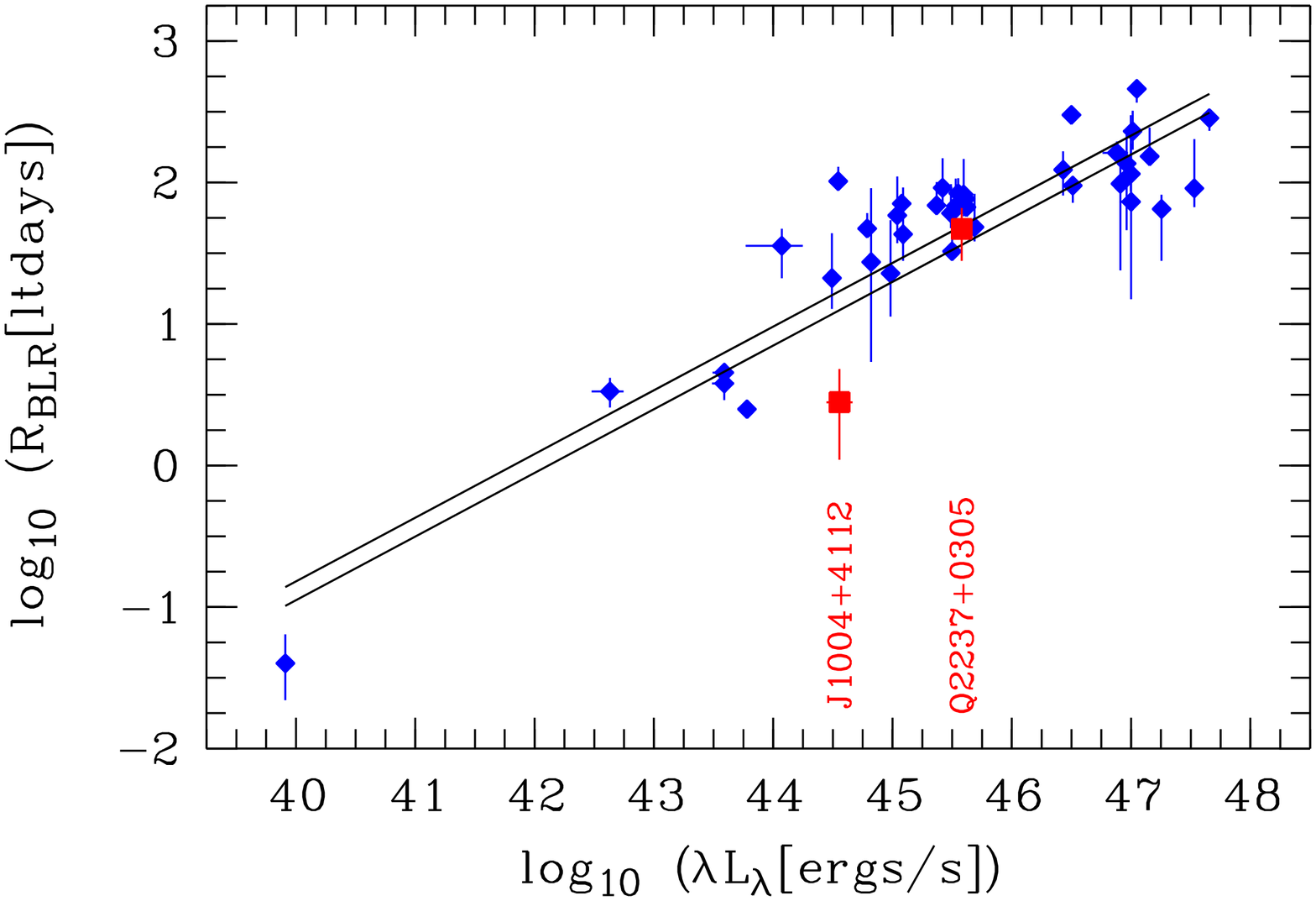}}
\caption{Relation between the radius of the \ion{C}{iv} BLR, that is, the rest-frame time lag from reverberation mapping and the continuum luminosity at 1350 \AA , with different fits superimposed as continuous lines (from \citealt{2021Kaspi}). The BLR half-light radii measured from our microlensing analysis of the quasars J1004+4112 and Q2237+0305 are superimposed in red.}
\label{fig:kaspir}
$ $ \\
\resizebox{\hsize}{!}{\includegraphics*{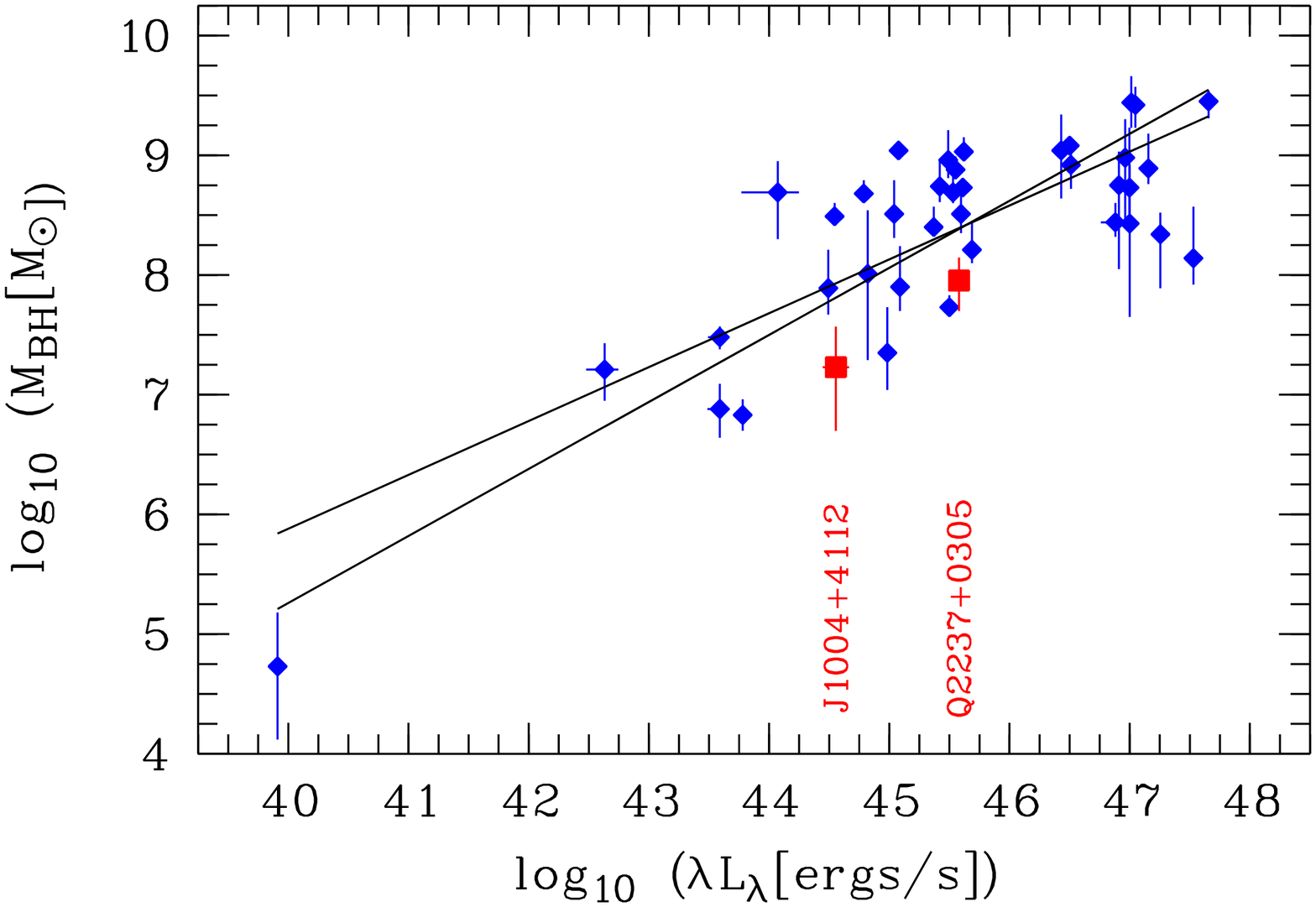}}
\caption{Relation between the black hole mass and the continuum luminosity at 1350 \AA , with different fits superimposed as continuous lines (from \citealt{2021Kaspi}). The black hole masses measured from our microlensing analysis of the quasars J1004+4112 and Q2237+0305 are superimposed in red.}
\label{fig:kaspim}
\end{figure}

\citet{2020Rakshit} reported the luminosity $\lambda L_{\lambda} (1350 \AA)$ for image A in January 2016. Divided by the macro-model magnification factor 29.7 \citep{2010Oguri}, it amounts to  $\lambda L_{\lambda} (1350 \AA) = 1.9 \times 10^{44}$ ergs s$^{-1}$. \citet{2020Popovic} measured this luminosity in February 2018, which, corrected for the macro-model magnification factor 29.7, amounts to $\lambda L_{\lambda} (1350 \AA)$ = $3.2 \times 10^{44}$ ergs s$^{-1}$. Since the object does not seem to be brighter in 2018 than in 2016 \citep{2022Munoz}, hereafter, we use the average value $\lambda L_{\lambda} (1350 \AA)$ = $3.6 \pm 0.9 \times 10^{44}$ ergs s$^{-1}$, corrected for the microlensing demagnification $\mu^{cont} \simeq 0.7$ that is likely to occur for the 2016-2018 data (Fig.~\ref{fig:indall}). With this luminosity, we derived the expected \ion{C}{iv} BLR radius $R_{\rm BLR}$ from the radius-luminosity ($R-L$) relations of \citet{2021Kaspi} based on reverberation mapping:  $R_{\rm BLR} = 15^{+15}_{-7}$ light days. The reverberation mapping BLR radius is the rest-frame time lag measured as the centroid of the cross-correlation function between the continuum and line light curves. It corresponds (in theory) to the the luminosity-weighted radius of the BLR \citep{1991Koratkar} and can thus be compared to the half-light radius derived from microlensing. Our estimate of the BLR size, $r_{1/2} = 2.8^{+2.0}_{-1.7}$ light-days, is smaller by a factor of 5. However, the errors are large: a radius of 2.8 light-days represents only a 1.7 $\sigma$ departure from the $R-L$ relation, which shows several outliers with comparable differences (Fig.~\ref{fig:kaspir}). Given that the range of BLR sizes involved in the $R-L$ relation covers four orders of magnitude, our BLR radius appears to be in reasonable agreement, as it is only marginally smaller than expected. The BLR radius estimated for Q2237+0305 \citep{2021Hutsemekers} is in excellent agreement with the $R-L$ relations, indicating that the microlensing values are not biased\footnote{We recomputed the \ion{C}{iv} half-light radius of Q2237+0305 using the method described in the present paper, that is using the full $\mu(v)$ profile. We found $r_{1/2} = 39^{+17}_{-25}$  light-days in excellent agreement with the value reported in \citet{2021Hutsemekers}.}. On the other hand, they depend on the microlens mass, which could differ from the average value of 0.3 M$_{\odot}$, especially in a system like J1004+4112, which is lensed by a cluster. For the microlensing BLR radius to precisely fit the $R-L$ relation, the microlens mass should be changed from the average value of 0.3 M$_{\odot}$ to 7.5 M$_{\odot}$.

\subsection{Black hole mass}
\label{sec:mass}

Assuming virial motion, the mass of the black hole can be estimated based on:
\begin{equation}
\label{eq:mbh}
M_{\rm BH} = f \, \frac{R_{\rm BLR} \, V^2}{G}
,\end{equation}
where $V$ is the velocity full width at half-maximum (FWHM) of the broad emission line, $f$ is the virial factor, and $G$ is the gravitational constant. For a thin disk, \citet{2008Decarli} give\footnote{The factor $f$ given in \citet{2008Decarli} is the square root of  the factor $f$ used in Eqs.~\ref{eq:mbh} and~\ref{eq:f}.}:
\begin{equation}
f = (4 \, \sin^2i)^{-1} \, .
\label{eq:f}
\end{equation}
For the broad \ion{C}{iv} line observed in the 2018 spectra of image B (unaffected by microlensing), we measured FWHM = 4900$\pm$300 km~s$^{-1}$. Using $i$ = 35$\pm$10\degr\ and $R_{\rm BLR}$ =  $r_{1/2} = 2.8^{+2.0}_{-1.7}$ light-days, we finally obtained $M_{\rm BH} = 1.0^{+1.1}_{-0.7}\times 10^7$ M$_{\odot}$. To correct the bias that affects the \ion{C}{iv}-based virial black hole masses, we used the prescriptions of \citet{2017Coatman}. With a \ion{C}{iv} blueshift of 350$\pm$50 km~s$^{-1}$ measured in the image B spectrum, the corrected black hole mass amounts to  $M_{\rm BH} = 1.7^{+2.0}_{-1.2}\times 10^7$ M$_{\odot}$. As the BLR radius, this black hole mass is a factor of 4-5 smaller than the value expected from the black hole mass-luminosity  ($M_{\rm BH}-L$) relations of \citet{2021Kaspi}, $M_{\rm BH} = 7.5^{+11.2}_{-4.2}\times 10^7$ M$_{\odot}$. It is nevertheless in agreement within the uncertainties, due to the large dispersion of the $M_{\rm BH}-L \,$ correlation (Fig.~\ref{fig:kaspim}). The black hole mass of Q2237+0305 derived in \citet{2021Hutsemekers} is in better agreement with the $M_{\rm BH}-L \,$ relation. 


\section{Conclusions}
\label{sec:conclusion}

Based on 15 spectra obtained from 2003 to 2018, we revisited the microlensing effect that distorts the \ion{C}{iv} broad emission line profile in the lensed quasar J1004+4112. We took advantage of recent measurements of the image macro-magnification ratios and of the fact that, at one epoch, image B was not microlensed, thus constituting a reference spectrum that allowed us to unambiguously characterize the microlensing effect observed in image A. After disentangling the microlensing in images A and B, we showed that the microlensing-induced line profile distortions in image~A, although variable, are remarkably similar over a period of 15 years. They are characterized by a strong magnification of the blue part of the line profile, a strong demagnification of the red part of the line profile, and a small to negligible demagnification of the line core. Microlensing also affects the \ion{C}{iv} line profile in image~B, but to a lesser extent.

We then compared the microlensing effect observed in image~A, characterized by the magnification profile of the \ion{C}{iv} emission line or by a set of four integrated indices, to simulated line-profile deformations that result from the selective magnification of BLR models by caustic networks. Our main conclusions are as follow.\ 
\begin{itemize}
\item The observed $\mu(v)$ magnification profile of \ion{C}{iv} in J1004+4112 can be reproduced with the simple BLR models we considered, without the need for more complex BLR features. The magnification pattern is dominated by the position of the BLR with respect to the caustic network -- and not by the velocity-dependent size of the BLR. The observation of such a microlensing effect over 15 years has been deemed plausible. 
\item The favored models for the \ion{C}{iv} BLR are either the Keplerian disk or the equatorial wind, depending on the orientation of the caustic map with respect to the BLR axis. The polar wind model is much less likely and can be rejected. In Q2237+0305, the most likely BLR model is that of a Keplerian disk \citep{2021Hutsemekers}.
\item We measured the \ion{C}{iv} BLR half-light radius as $r_{1/2} = 2.8^{+2.0}_{-1.7}$ light-days. This result is marginally lower than the BLR radius expected from empirical $R-L$ relations, while the \ion{C}{iv} BLR half-light radius measured in Q2237+0305 \citep{2021Hutsemekers} is in excellent agreement with these relations. More measurements are needed to compare the BLR sizes estimated from reverberation mapping with those derived from microlensing. 
\end{itemize}

Finally, it is interesting to note that similar conclusions on the preferred BLR models can be reached using either the four indices or the full $\mu(v)$ magnification profile. While both can be used for single-epoch microlensing analyses, this result validates the use of indices, which can ultimately prove more convenient for interpreting time series of line-profile deformations.

\begin{acknowledgements}
D.H. and D.S acknowledge support from the Fonds de la Recherche Scientifique - FNRS (Belgium) under grants PDR~T.0116.21 and No 4.4503.19.
\end{acknowledgements}

\bibliographystyle{aa}
\bibliography{aa45490}

\end{document}